\documentclass[12pt]{article}

\usepackage[utf8]{inputenc}

\usepackage[margin=1in]{geometry}
\usepackage{setspace}

\usepackage{amsmath, amsthm, amssymb, amsfonts}
\usepackage{bbm}
\usepackage{bigints}

\usepackage{graphicx}
\graphicspath{{Pictures/}}
\usepackage{caption}
\usepackage{subcaption}
\usepackage{float}

\usepackage{booktabs}
\usepackage{multirow}

\usepackage{tikz}
\makeatletter
\IfFileExists{tikzlibrarybayesnet.code.tex}{\usetikzlibrary{bayesnet}}{}
\makeatother

\usepackage[toc,page]{appendix}

\usepackage{natbib}
\usepackage{url}

\usepackage{xcolor}
\usepackage{enumitem}
\usepackage{enumerate}
\usepackage{authblk}
\usepackage{framed}
\usepackage[normalem]{ulem}
\usepackage{titlesec}
\usepackage{comment}
\usepackage{pdfpages}
\usepackage{epstopdf}

\usepackage[colorlinks,citecolor=blue,urlcolor=blue]{hyperref}

\usepackage{amsmath}
\usepackage{amsthm}
\usepackage{amssymb}
\usepackage{graphicx}
\usepackage{color}
\usepackage{enumerate}
\usepackage{natbib}
\usepackage{url} 
\usepackage{multirow}
\usepackage{lineno}
\usepackage{makecell}
\usepackage{hyperref}
\usepackage{algorithm}
\usepackage{algpseudocode}

\newcommand{\bm}{\boldsymbol}

\def\0{\boldsymbol 0}

\theoremstyle{definition}
\newtheorem{theorem}{Theorem}

\newtheorem{Aass}{}
\newtheorem{Bass}{}
\title{\bf Green's matching: an efficient approach to parameter estimation in complex dynamic systems}
  \author{Jianbin Tan \thanks{Guoyu Zhang is the co-first author.}\\
  International Institute of Finance, School of Management, \\
   University of Science and Technology of China\\
   and \\
    School of Mathematics, Sun Yat-sen University\\
    Guoyu Zhang\\
   Department of Probability and Statistics, \\
   School of Mathematical Sciences, Center for
Statistical Science, Peking University\\
Xueqin Wang\thanks{Corresponding author.}\\
   International Institute of Finance, School of Management, \\
   University of Science and Technology of China\\
   Hui Huang$^{\dagger}$ \\
   Center for Applied Statistics, School of Statistics, Renmin University of China\\
    and\\
    Fang Yao$^{\dagger}$\\
    Department of Probability and Statistics, \\
   School of Mathematical Sciences, Center for
Statistical Science, Peking University\\
    }

\date{}

\begin{document}
\maketitle

\begin{abstract}
Parameters of differential equations are essential for characterizing the intrinsic behaviors of dynamic systems. 
Numerous methods for estimating parameters in dynamic systems are computationally and/or statistically inadequate, especially for complex systems with general-order differential operators, such as motion dynamics.
This article presents Green's matching, a computationally tractable and statistically efficient two-step method, which only requires approximating trajectories in dynamic systems but not their derivatives, due to the inversion of differential operators via Green's function.
This approach yields a statistically optimal guarantee for parameter estimation in general-order equations, a feature not shared by existing methods, and provides an efficient framework for broad statistical inference in complex dynamic systems.
\end{abstract}


\setstretch{1.5} 
\section{Introduction}\label{introduction}
Dynamic systems have found broad applications in modeling the evolution of systems in various domains, such as the physical motion of objects \citep{spring-mass}, the evolutionary paths of genes \citep{wu2014sparse,chen2017network}, and the transmission dynamics of epidemics \citep{tian2021effects,tan2023age}.
Among these systems, their dynamic behaviors are usually characterized by differential equations:
\begin{equation}\label{dyn}
\mathcal{D}_i^K X_i(t) = {f}_i(\bm{X}(t), t; \bm{\beta}), \quad t \in [0, C]\ \text{and}\ i=1,\cdots,p,
\end{equation}
where $X_i(t)$ and $\bm{X}(t)=\left(X_{1}(t),\cdots,X_{p}(t)\right)^T$ are temporal trajectories of the target system, $\mathcal{D}_i^K$ is an order-$K$ differential operator, and $f_i(\cdot;\bm{\beta})$ is a driving function indexed by parameters $\bm{\beta}$. Typically, $\mathcal{D}_i^K$ and $f_i(\cdot;\bm{\beta})$ are two important ingredients that describe the intrinsic behaviors and interdependence among system variables.
The former determines the derivatives of curves (e.g., velocity or acceleration) driven by the latter, and the form of $f_i(\cdot;\bm{\beta})$ is usually grounded in specific fields, with $\bm{\beta}$ to be estimated from empirical data.
Parameter estimation in dynamic systems facilitates the recognition of evolution patterns from observed data, thereby improving our understanding of system variables and enabling predictions of future dynamics.

There is a vast literature on parameter estimation in dynamic systems, which can be broadly summarized into two frameworks. The first framework is known as ``nonlinear least squares,'' which estimates the parameters by fitting the forward solution of \eqref{dyn} to the observed data \citep{ramsay2017dynamic}. However, when $f_i(\cdot;\bm{\beta})$ in \eqref{dyn} is a nonlinear function, such a trajectory-matching approach requires repeated equation solving in its estimation steps, leading to substantial computational demands for parameter estimation.
Alternatively, the second framework utilizes smoothing approaches to approximate system trajectories and their derivatives for parameter estimation. This framework includes a family of collocation methods following a ``first smoothing, then estimation'' two-step strategy \citep{varah1982spline, brunel2008parameter, Shota, Itai}, where the accuracy of the estimated parameters depends on the quality of the initial smoothing.
To improve performance, iterative steps can be incorporated into collocation methods for a better approximation of curves \citep{ramsay2007parameter, niu2016fast, yang2021inference}. Nonetheless, these approaches may incur additional computational costs in their estimation steps.
In many cases, especially for complex dynamic systems, the original two-step strategy receives more attention in statistical inference \citep{wu2014sparse, chen2017network, zhang2020bayesian, dai2021kernel} due to its simple implementation and efficient computation.

Gradient matching \citep{brunel2008parameter, Shota} and integral matching \citep{Itai, ramsay2017dynamic} are two important collocation methods, which are not only employed for parameter estimation but can also be extended to more complicated tasks, such as equation discovery \citep{brunton2016discovering, champion2019data} or causal network inference \citep{wu2014sparse, chen2017network, dai2021kernel} for dynamic systems.
In general, these collocation methods focus on model fitting by matching the unknown parameters with trajectories in equations, where the trajectories are either fully observed or substituted by smooth approximations from discrete noisy data.
Such an equation-matching approach avoids time-consuming equation solving in parameter estimation, thereby providing significant computational benefits for many statistical tasks in dynamic systems.
In addition to the computational advantages, the statistical properties of gradient matching and integral matching have also been discussed in the literature \citep{brunel2008parameter, Shota, Itai}. These studies demonstrated that both methods achieve root-$n$ consistency for parameter estimation in first-order systems.

Despite their importance, these two methods mainly focus on systems with known first-order differential operators rather than cases with unknown higher-order differential operators. The latter setting is more general and covers a broad range of systems with general-order evolution patterns, such as motion dynamics.
For these complex systems, direct extensions of gradient matching and integral matching may require pre-smoothing of curve derivatives. Nonetheless, pre-smoothing derivatives can be inaccurate, or even inappropriate, when the actual curves are not smooth (e.g., the acceleration of a motion trajectory could be discontinuous due to a sudden change of force). This can lead to significant biases and poor theoretical behavior for parameter estimation.
Targeting general-order dynamic systems, estimating parameters and differential operators is essential but rarely discussed in the literature. There is an urgent need to develop a computationally tractable and statistically efficient approach for these purposes.

In this article, we adopt the ``first smoothing, then estimation'' two-step strategy for parameter estimation in general-order dynamic systems. 
Our study focuses on cases where the trajectories of dynamic systems are discretely and noisily observed, and we apply local polynomial regression \citep{fan2018local} to pre-smooth the trajectories from discrete noisy data. 
The proposed method is called \textbf{Green's matching}, which utilizes Green's functions of differential operators \citep{duffy2015green} to incorporate pre-smoothing of trajectories into parameter estimation.
As an important mathematical tool, Green's function is a kernel function that is considered an inversion operation of a differential operator. It finds diverse applications ranging from approximating forward solutions of differential equations \citep{wahba1973class, duffy2015green} to statistical inference for differential-equation-based models \citep{heckman2000penalized, boullé2022learning, stepaniants2023learning}.
However, these studies do not focus on estimating parameters and differential operators simultaneously using discretely and noisily observed trajectory data. Our proposed method fills this gap, providing an efficient approach for statistical inference in general-order systems.

Overall, the superiority of Green's matching lies in both statistical and computational aspects.
First, our method utilizes Green's functions to invert the general-order differential operators in \eqref{dyn}, thereby eliminating the need for pre-smoothing curve derivatives within parameter estimation.
This appealing feature reduces the dependence on pre-smoothing for parameter estimation in general-order systems, compared to both gradient matching and integral matching.
Furthermore, if $f_i(\cdot;\bm{\beta})$ in \eqref{dyn} satisfies a general separability condition, Green's matching degenerates to the conventional least squares fitting, greatly reducing the computational effort required for general-order systems. 

Under mild conditions, we establish root-$n$ consistency and asymptotic normality for Green's matching. Moreover, we extend the ideas of gradient matching and integral matching to general-order dynamic systems and investigate their asymptotic properties. The estimation biases and convergence rates of these methods are both justified and discussed. As shown, the estimation bias increases as more pre-smoothing is used within parameter estimation. Furthermore, we reveal that gradient matching and integral matching may not achieve root-$n$ consistency as the differential order of dynamic systems increases. In contrast, Green's matching always attains this optimal rate for parameter estimation.

The rest of this article is organized as follows. In Section~\ref{Inma}, we first propose Green's matching for parameter estimation in general-order dynamic systems. The statistical properties of Green's matching and other two-step methods are then established in Section~\ref{prope}.  
Moreover, we conduct extensive simulation studies in Section~\ref{sim_dat_bo} to investigate the estimation behaviors of the two-step approaches and other competitive methods \citep{ramsay2007parameter, yang2021inference}. 
Finally, we apply Green's matching to equation discovery for a nonlinear pendulum in Section~\ref{ED}, and present conclusions and discussions in Section~\ref{dis}. The code, data, and proofs for this article can be found at \href{https://github.com/Tan-jianbin/Statistical-Inference-in-General-order-Dynamic-Systems}{https://github.com/Tan-jianbin/Statistical-Inference-in-General-order-Dynamic-Systems}.

\section{Green's matching}\label{Inma}
We first introduce an observational model for dynamic systems. Let $Y_{i}(t_j)$ be a measurement of $X_i(t_j)$
\begin{equation}\label{mea_model}
    {Y}_{i}(t_j)={X}_{i}(t_j)+\varepsilon_{i}(t_j),
\end{equation}
for $i=1,\cdots,p$, $j=1,\cdots,n$, and  $t_j\in [0,C]$,
where $\varepsilon_{i}(t_j)$s are mean-zero white noises, $X_i(t_j)$ is the mean of $Y_{i}(t_j)$ conditioning on the time point $t_j$, and $X_i(t)$, $i=1,\cdots,n$, are the smooth curves over $t\in[0,C]$ satisfying the differential equations \eqref{dyn}. We assume that $\mathcal{D}_i^K$ in \eqref{dyn} is a linear differential operator of order $K$
\begin{equation}\label{DO}
    \mathcal{D}_i^K=D^K+\sum_{k=0}^{K-1}\omega_{ik}D^k,
\end{equation}
where $\bm{\omega}:=(\omega_{ik})_{i\leq p, k\leq K}$ are the coefficients of $\mathcal{D}_i^K$s, and $D^k$ is the differential operation $\frac{\mathrm{d}^k}{\mathrm{d}t^k}$. 
Since the differential order can usually be determined based on prior knowledge of dynamic systems, we assume that $K$ is fixed in our analysis.
Under this framework, $X_i(\cdot)$s are determined by the parameters $\bm{\beta}$, differential operators $\mathcal{D}_i^K$s, and the initial conditions $\left\{D^kX_{i}(0);k=0,\cdots, K-1,i=1,\cdots,p\right\}$. 
In practice, the differential operators and initial conditions, like the parameters, are usually unknown, especially when $K>1$. Denote $\Omega_{\text{parameter}}$ as the parameter space of a parameter.
Our goal is to estimate $\bm{\beta}\in \Omega_{\bm{\beta}}$ and $\bm{\omega}\in \Omega_{\bm{\omega}}$ based on $\{Y_{i}(t_j);i=1,\cdots,p,j=1,\cdots,n\}$ without providing initial conditions.

\subsection{Local polynomial regression}\label{pres}

We employ local polynomial regressions \citep{fan2018local} to pre-smooth the trajectories of systems from the discrete noisy data.
Given $k\leq K$ and $t\in [0,C]$, we estimate $D^kX_i(t)$ by $\widehat{D^kX}_i(t):=a_k$, among $a_k$ minimizes
\begin{equation}\label{local_est_1}
    \sum_{j=1}^nW\left(\frac{t_j-t}{h_{k}}\right)\left\{Y_{i}(t_j)-\sum_{v=0}^{m_k}a_v\frac{(t_j-t)^v}{v!}\right\}^2,
\end{equation}
where $m_k\geq k$ is the order of the polynomials, $W(\cdot)$ is a function that assigns weights, and $h_{k}$ is a bandwidth to control the weighted smoothing for the $k^{th}$ derivative. 
The value of $m_k$ influences the asymptotic properties of $\widehat{D^kX}_i(\cdot)$ in relation to the smoothness of $X_i(\cdot)$ \citep{fan1994}. 
In many cases, we may only know that $X_i(\cdot)$ is at least $K$-times differentiable, whereas its higher-order smoothness can be further confirmed by the driving function in \eqref{dyn}.  For simplicity, we set that $m_k=k+1$ as recommended in \cite{fan1994}.
Moreover, we utilize the Epanechnikov weight function $W(t)=\frac{3}{4}\max(1-t^2,0)$ to conduct the local polynomial fitting due to its asymptotic optimality \citep{fan1997}.
Further sensitivity analysis for the selection of weight functions is given in Part 3.2 in Supplementary Materials.

\begin{figure}[h]
\centering
\includegraphics[scale = 0.73]{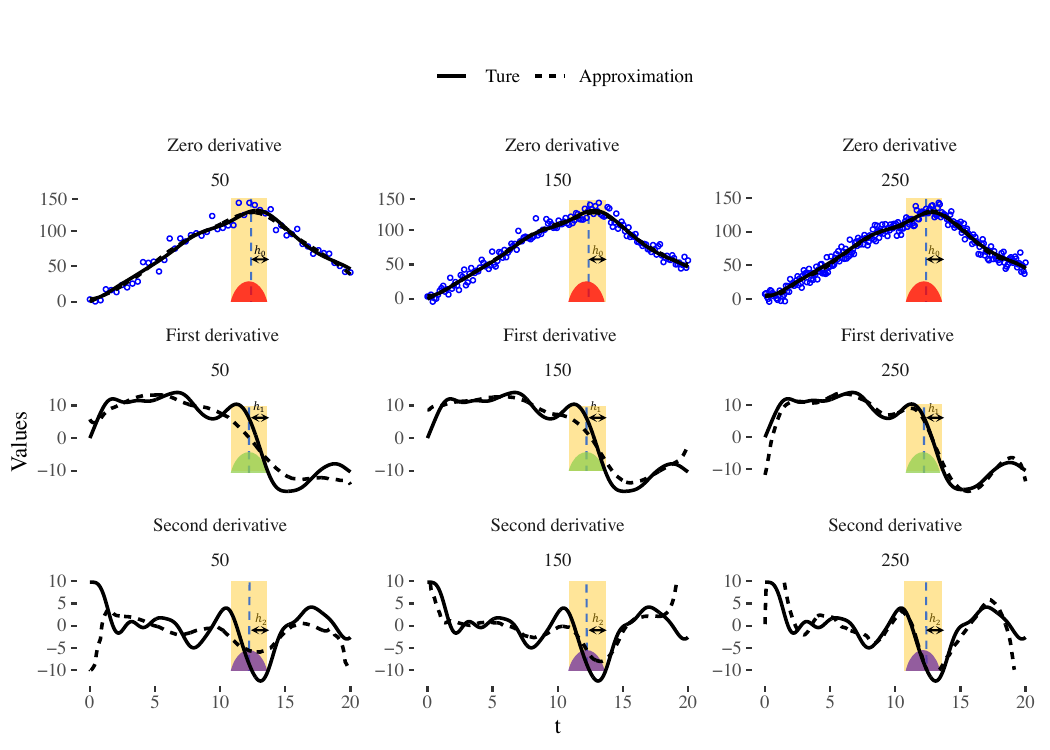}
\caption{The estimated derivatives for a curve (in a spring-mass system) by local polynomial regressions, where the estimation is performed based on the discrete noisy data (blue circles) obtained from different sample sizes $n$ (subtitles). The red, green, and purple regions represent the weights assigned to the observations, controlled by the bandwidths $h_0$, $h_1$, and $h_2$, respectively. For more information about the spring-mass system and the bandwidth selection scheme, please refer to Section \ref{sim_1} and Part 1.1 in the Supplementary Materials, respectively.}\label{local}
\end{figure}

With some conditions and a suitably chosen bandwidth $h_k$, $\widehat{D^kX}_i(\cdot)$ serves as a consistent estimator of ${D^kX}_i(\cdot)$ as $n\rightarrow\infty$ \citep{fan1994}, although its statistical convergence rate may slow down as $k$ increases \citep{stone1982optimal}.
To illustrate this phenomenon, we present several examples of local polynomial regressions in Figure \ref{local}. We observe that the estimated zero derivatives of curves accurately match their real shapes for different sample sizes $n$. However, the estimation accuracy diminishes for the higher-order derivatives. This discrepancy is likely due to the difficulties in capturing high-order effects of curves from limited noisy data.

\subsection{Equation matching for dynamic systems}

In this subsection, we employ the pre-smoothed curves for parameter estimation in general-order dynamic systems. In the following, we denote $\bm{X}(t)$ and $D^k\bm{X}(t)$ as $\bm{X}$ and $D^k\bm{X}$ to omit their functional nature.
To be general, let $\mathcal{T}_{t,\bm{\beta},\bm{\omega}}$ be an operator from $\bm{X}$, $D\bm{X}$, $\cdots$, $D^K\bm{X}$ to $\mathbb{R}^p$ s.t.
\begin{equation}\label{EE}
    \mathcal{T}_{t,\bm{\beta},\bm{\omega}}(\bm{X}, D\bm{X}, \cdots,D^K\bm{X})=\bm{0},\ \forall\ t\in [0,C],
\end{equation}
where $\bm{\beta}$ and $\bm{\omega}$ are taken as their true values in \eqref{dyn}, determining the relations among $\bm{X}$, $D\bm{X}$, $\cdots$, $D^K\bm{X}$.
When $\mathcal{T}_{t,\bm{\beta},\bm{\omega}}(\cdot)$ does exists, we can plug the pre-smoothed $\bm{{X}}$ and ${D^k\bm{X}}$ (denoted as $\bm{\hat{X}}$ and $\widehat{D^k\bm{X}}$) into \eqref{EE} and obtain
\begin{equation}\label{me_dyn}
    \mathcal{T}_{t,\bm{\beta},\bm{\omega}}(\bm{\hat{X}}, \widehat{D\bm{X}},  \cdots,\widehat{D^K\bm{X}})\approx\bm{0}, \forall t\in [0,C].
\end{equation}
As the unknown identities in \eqref{me_dyn} only consist of $\bm{\beta}$ and $\bm{\omega}$, we can utilize these equations, which are called the matching equations, to construct estimators for the unknown parameters.
Note that there are infinitely many equations in \eqref{me_dyn}.
To combine information from these equations, we employ an integration strategy to estimate $\bm{\beta}$ and $\bm{\omega}$ by minimizing them from
\begin{eqnarray*}
\int_{0}^C\left\|\mathcal{T}_{t,\bm{\beta},\bm{\omega}}(\bm{\hat{X}}, \widehat{D\bm{X}},  \cdots,\widehat{D^K\bm{X}})\right\|^2 w(t)\ \mathrm{d}t,
\end{eqnarray*}
where $||\cdot||$ is the Euclidean norm and $w(\cdot)$ is a known function on $[0,C]$ to assign weights for different matching equations. 
The fitting procedure above is called equation matching, which serves as a general framework to avoid equation solving for parameter estimation in dynamic systems. We here review two important equation-matching methods: gradient matching and integral matching.

Let $\mathcal{T}^i_{t,\bm{\beta},\bm{\omega}}(\cdot)$ be the $i^{th}$ component of $\mathcal{T}_{t,\bm{\beta},\bm{\omega}}(\cdot)$. We first assume $K=1$, where $\mathcal{D}_i^K$ degenerates to the first-order operation $D$, and we simplify $\mathcal{T}^i_{t,\bm{\beta},\bm{\omega}}(\cdot)$ as $\mathcal{T}^i_{t,\bm{\beta}}(\cdot)$. 
By \eqref{dyn}, we can directly establish a matching equation by $\mathcal{T}^i_{t,\bm{\beta}}(\bm{X},D\bm{X}) = D{X}_i(t) - {f}_i(\bm{{X}}(t),t;\bm{\beta})$. This can be realized as fitting the unknown parameters to the gradients of curves. Therefore, the resulting estimator is called gradient matching \citep{brunel2008parameter, Shota}. It should be noted that the gradient matching incorporates inaccurately pre-smoothed derivatives for estimating parameters.
Alternatively, one can construct another matching equation using the Newton-Leibniz formula of \eqref{dyn}, or equivalently, we set $\mathcal{T}^i_{t,\bm{\beta},\alpha_i}(\bm{X})={X}_i(t) - {\alpha}_i - \int_0^{t}{f}_i({\bm{X}}(s),s;\bm{\beta})\ \mathrm{d}s$, where $\alpha_i$ is an unknown parameter determined by the initial conditions of \eqref{dyn}. The resulting estimator is known as integral matching \citep{ramsay2017dynamic}, which removes $D\bm{X}$ from the matching equations by integrating \eqref{dyn}. Consequently, it avoids using $\widehat{D\bm{X}}$ within parameter estimation.

We now generalize the above ideas to high-order systems. For $K>1$, we define ${F}_i(t;\bm{\beta},\bm{\omega})= {f}_i(\bm{{X}}(t),t;\bm{\beta})-\sum_{k=0}^{K-1}\omega_{ik}{D^k{X}}_i(t)$ and rewrite \eqref{dyn} as 
$D^KX_i(t)={F}_i(t;\bm{\beta},\bm{\omega})$.
Therefore, we can similarly construct matching equations
\begin{eqnarray}
\mathcal{T}^{i,K}_{t,\bm{\beta},\bm{\omega}}(\bm{X}, D\bm{X}, \cdots,D^K\bm{X})&=&
{D^K{X}}_i(t)- {F}_i(t;\bm{\beta},\bm{\omega}),\label{gra_mat}\\
\mathcal{T}^{i,K-1}_{t,\bm{\beta},\bm{\omega},\alpha_i}(\bm{X}, D\bm{X}, \cdots,D^{K-1}\bm{X})&=&D^{K-1}X_i(t) -\alpha_{i}-\int_0^t {F}_i(s;\bm{\beta},\bm{\omega})
\ \mathrm{d}s.\label{intm}
\end{eqnarray}
We call the equation matching based on $\mathcal{T}^{i,K}_{t,\bm{\beta},\bm{\omega}}(\cdot)$ or $\mathcal{T}^{i,K-1}_{t,\bm{\beta},\bm{\omega},\alpha_i}(\cdot)$ as the order $K$ or $(K-1)$ gradient matching, respectively, as both methods fit the parameters to the gradients of curves.
Note that via the incorporation of integration, the order $(K-1)$ gradient matching reduces one differential order compared to the order $K$ gradient matching. Nonetheless, these approaches both involve pre-smoothing of high-order derivatives for general-order systems, thereby imposing additional pre-smoothing errors. This is even inappropriate when the smooth high-order derivative does not exist, as shown in Section \ref{sim_1}, 
Therefore, we require a new approach capable of eliminating all high-order derivatives in matching equations for high-order systems.

\subsection{Generalized integral matching by Green's function}
In this subsection, we propose an efficient approach called Green's matching for parameter estimation, which removes the pre-smoothed derivatives via Green's functions of differential operators \citep{duffy2015green}.
In general, let $\mathcal{D}$ be a differential operator of functions on $[0, C]$. We define Green's function of $\mathcal{D}$ as a kernel function $G(t,s)$ on $[0, C]^2$ satisfying $\mathcal{D}G(t,s)=\delta(t-s)$, where $\mathcal{D}$ is applied to the variable $t$, and $\delta(\cdot)$ is a Dirac delta function at zero. Given $G(t,s)$ of $\mathcal{D}$ and any functions $X(t)$ and $f(t)$ s.t. $\mathcal{D}X(t)=f(t)$, we have $X(t)=g(t)+\int_{0}^C G(t,s)f(s)\ \mathrm{d}s$, where $g(\cdot)$ is any function in the kernel space of $\mathcal{D}$, denoted as $\operatorname{Ker}(\mathcal{D}) := \left\{g(\cdot);\mathcal{D}g(t)=0,\forall t\in [0,C]\right\}$.
In theory, $G(t,s)$ is considered an inverse operation for the operator $\mathcal{D}$, which would be used to reduce the differential order of the matching equations.

We first show that the Green's function of $\mathcal{D}^K_i$, denoted as $G_{i}^K(t,s)$, can be constructed as follows. First note that $\operatorname{Ker}(\mathcal{D}^K_i)$ is a $K$-dimensional function space. Denote $\bm{\psi}_i(t)\in \mathbb{R}^K$ as the basis functions of $\operatorname{Ker}(\mathcal{D}^K_i)$, which can be constructed as
\begin{eqnarray}\label{basis}
\bm{\psi}_i(t)=\left([\textbf{e}^{t \bm{A}(\bm{\omega}_i)}]_{1,1},\cdots,[\textbf{e}^{t \bm{A}(\bm{\omega}_i)}]_{1,K}\right)^T,
\end{eqnarray}
where  $\textbf{e}^{\mathbf{M}}$ is the matrix exponential of a matrix $\mathbf{M}$, $[\ ]_{m,m^{\prime}}$ extracts the $(m,m^{\prime})^{th}$ element of a matrix, $\bm{\omega}_i=(\omega_{i0},\cdots,\omega_{i(K-1)})^T$, and $\bm{A}(\bm{\omega}_i)$ is a $K\times K$ matrix with its $K^{th}$ row being $\bm{\omega}_i^T$ and $[\bm{A}(\bm{\omega}_i)]_{k,k^{\prime}}=\mathbb{I}(k=k^{\prime}-1)$ for $k\leq K-1$ and $k^{\prime}\leq K$, among $\mathbb{I}(\cdot)$ is an indicator function.
Given $\bm{\psi}_i(t)$, $G_i^K(t,s)$ can be constructed as
\begin{eqnarray}\label{green_function_9}
    G_i^K(t,s)=\left[\textbf{e}^{(t-s)\bm{A}(\bm{\omega_i})}\right]_{1,K}\mathbb{I}(t\geq s).
\end{eqnarray}
The proof of \eqref{basis} and \eqref{green_function_9} can be found in Part 1.2 in Supplementary materials. When $\bm{\omega}_i=\bm{0}$, the above equations indicate that 
$\bm{\psi}_{i}(t)=\left(1,t,\cdots,\frac{t^{K-1}}{(K-1)!}\right)^T$ and
$G_i^K(t,s)=\frac{(t-s)^{K-1}}{(K-1)!}\mathbb{I}(t\geq s)$.

Given the Green's function $G_{i}^K(t,s)$ of $\mathcal{D}^K_i$, we can then transform the differential equation \eqref{dyn} into the following integral equation
\begin{equation}\label{green_inverse_2}
{X}_{i}(t)=\{\bm{\psi}_i(t)\}^T\bm{\alpha}_{i}+\int_{0}^CG_i^K(t,s)f_i(\bm{X}(s),s;\bm{\beta})\ \mathrm{d}s, 
\end{equation}
where $\bm{\alpha}_i\in \mathbb{R}^K$. 
Note that $G_i^K(t,s)$ in \eqref{green_inverse_2} may have a highly nonlinear relation with $\bm{\omega}_i$ due to \eqref{green_function_9}.
To simplify \eqref{green_inverse_2} when $\bm{\omega}_i$s are unknown, we utilize the linear structure of differential operators and instead apply the Green's function of $D^K$ to the both sides of the equation \eqref{dyn}. As a result, we obtain
\begin{equation}\label{green_inverse}
{X}_{i}(t)+\sum_{k=0}^{K-1}\omega_{ik}\int_{0}^C {G}_{K-k}(t,s){X}_i(s)\ \mathrm{d}s=\{\bm{\psi}(t)\}^T\bm{{\alpha}}_{i}+\int_{0}^CG_K(t,s)f_i(\bm{X}(s),s;\bm{\beta})\ \mathrm{d}s,
\end{equation}
where $\bm{\psi}(t)=\left(1,t,\cdots,\frac{t^{K-1}}{(K-1)!}\right)^T$ is the basis function of $\operatorname{Ker}(D^K)$, and $G_k(t,s)=\frac{(t-s)^{k-1}}{(k-1)!}\mathbb{I}(t\geq s)$ is the Green's function of $D^k$ as constructed by \eqref{green_function_9}. Different form \eqref{green_inverse_2}, the left side of \eqref{green_inverse} has a linear relation with $\bm{\omega}_i$, which largely simplifies the following estimation for $\bm{\omega}$ when the differential operator is unknown. 

Define $\mathcal{T}^{i,\text{Gree}}_{t,\bm{\beta},\bm{\omega},\bm{\alpha}_i}(\bm{X}):={X}_{i}(t)+\sum_{k=0}^{K-1}\omega_{ik}\int_{0}^C {G}_{K-k}(t,s) {X}_i(s)\ \mathrm{d}s-\{\bm{\psi}(t)\}^T\bm{{\alpha}}_{i}-\int_{0}^CG_K(t,s)$ $f_i(\bm{X}(s),s;\bm{\beta})\ \mathrm{d}s$ based on \eqref{green_inverse}. We utilize $\bm{\hat{{X}}}$ to construct estimators as 
\begin{eqnarray}
\min_{\bm{{\alpha}}\in\Omega_{\bm{{\alpha}}},\bm{\beta}\in\Omega_{\bm{\beta}},\bm{\omega}\in\Omega_{\bm{\omega}}}&\int_{0}^C\sum_{i=1}^p\left[\mathcal{T}^{i,\text{Gree}}_{t,\bm{\beta},\bm{\omega},\bm{\alpha}_i}(\bm{\hat{X}})\right]^2w(t)\ \mathrm{d}t,\label{gre_22}
\end{eqnarray}
where $\bm{{\alpha}}=\left(\bm{{\alpha}}_1^T,\cdots,\bm{{\alpha}}_p^T\right)^T$. Similarly, we can define $\mathcal{T}^{i,\text{Gree}}_{t,\bm{\beta},\bm{\alpha}_i}(\bm{X}):={X}_{i}(t)-\{\bm{\psi}_i(t)\}^T\bm{\alpha}_{i}-\int_{0}^CG_i^K(t,s)f_i(\bm{X}(s),s;\bm{\beta})\ \mathrm{d}s$ based on \eqref{green_inverse_2} when $\bm{\omega}$ in $\mathcal{D}_i^K$s are known, and conduct the estimation 
\begin{eqnarray}   \min_{\bm{{\alpha}}\in\Omega_{\bm{{\alpha}}},\bm{\beta}\in\Omega_{\bm{\beta}}}\int_{0}^C\sum_{i=1}^p\left[\mathcal{T}^{i,\text{Gree}}_{t,\bm{\beta},\bm{\alpha}_i}(\bm{\hat{X}})\right]^2w(t)\ \mathrm{d}t.\label{gre_11}
\end{eqnarray}
We call \eqref{gre_22} or \eqref{gre_11} as Green's matching.
Like gradient matching and integral matching, Green's matching does not assume known initial conditions for parameter estimation. 
Besides, it's worth noting that Green's matching reduces to integral matching when $K=1$. Thus, our method can serves as a high-order extension of integral matching, effectively eliminating the need for pre-smoothed derivatives in parameter estimation, whether the differential operator is known or not.

Notice that the Green's function $G_k(t,s)$ of ${D}^k$ is not unique, since all the function constructed as $G_k(t,s) + f_k(t,s)$, $\forall f_k(\cdot,s)\in \operatorname{Ker}({D}^k)$, is a valid Green's function of ${D}^k$.
Nonetheless, we can show that the estimated $\bm{\beta}$ and $\bm{\omega}$ by Green's matching are uniquely determined. In detail, 
since $\operatorname{Ker}({D}^k)\subset \operatorname{Ker}({D}^K)$ for $k\leq K$, let $\tilde{G}_k(t,s):=G_k(t,s) + f_k(t,s)$ be another Green's function of ${D}^k$, where $f_k(t,s)=\{\bm{\psi}(t)\}^T\bm{{u}}_k(s)$ for any function $\bm{{u}}_k(s)\in \mathbb{R}^K$. Given this,  $\mathcal{T}^{i,\text{Gree}}_{t,\bm{\beta},\bm{\omega},\bm{\alpha}_i}(\bm{X})$ in \eqref{gre_22} can also be expressed as 
$\mathcal{T}^{i,\text{Gree}^{\prime}}_{t,\bm{\beta},\bm{\omega},\tilde{\bm{\alpha}}_{i}}(\bm{X}):= {X}_{i}(t)+\sum_{k=0}^{K-1}\omega_{ik}\int_{0}^C \tilde{G}_{K-k}(t,s) {X}_i(s)\ \mathrm{d}s-\{\bm{\psi}(t)\}^T\tilde{\bm{\alpha}}_{i}-\int_{0}^C\tilde{G}_K(t,s)f_i(\bm{X}(s),s;\bm{\beta})\ \mathrm{d}s$,
where
\begin{eqnarray*}
    \tilde{\bm{\alpha}}_{i}=\bm{\alpha}_{i}-\int_{0}^C\bm{{u}}_{K}(s)f_i(\bm{X}(s),s;\bm{\beta}) \mathrm{d}s+\sum_{k=0}^{K-1}\omega_{ik}\int_{0}^C \bm{{u}}_{k}(s)  {X}_i(s) \mathrm{d}s.
\end{eqnarray*}
As a result, $\mathcal{T}^{i,\text{Gree}}_{t,\bm{\beta},\bm{\omega}_i,\bm{\alpha}_i}(\bm{\hat{X}})=\mathcal{T}^{i,\text{Gree}^{\prime}}_{t,\bm{\beta},\bm{\omega}_i,\tilde{\bm{\alpha}}_{i}}(\bm{\hat{X}})$ for all $\bm{\beta}$, $\bm{\omega}_i$, and $t$, implying that different choices of Green's functions in \eqref{gre_22} only result in a shift for the minimizer of $\bm{\alpha}$ and does not alter the minimizer of $\bm{\beta}$ and $\bm{\omega}$. Therefore, it is safe to employ the Green's function constructed by \eqref{green_function_9} for Green's matching.

\subsection{Optimization scheme for Green's matching}\label{op_gree_1} 

We focus on the optimization in (\ref{gre_22}) with $w(\cdot)$  being positive on a finite time grid $\{t_h\in [0,C];h=1,\cdots, H\}$, where $H$ is sufficiently large. 
Given the pre-smoothing of the trajectories $\hat{\bm{X}}$, we implement Green's matching as
\begin{equation}\label{sim_lik}
\min_{\bm{\beta}\in\Omega_{\bm{\beta}}}\sum_{i=1}^p\left\{\min_{{\bm{\nu}}_{i}\in\Omega_{\bm{\nu}_i}}
\bigg\|\bm{Z}_i-\bm{\Psi}_i\bm{\nu}_i -\bm{\zeta}_i(\bm{\beta})\right\|^2\bigg\},
\end{equation}
where 
$\bm{Z}_i=(\sqrt{w(t_1)}\hat{X}_{i}(t_1),\cdots,\sqrt{w(t_H)}\hat{X}_{i}(t_H))^T$,
$\bm{\Psi}_i=(\sqrt{w(t_1)}\bm{a}_i(t_1),\cdots,\sqrt{w(t_H)}\bm{a}_i(t_H))^T$ with 
$\bm{a}_i(t)=(\{\bm{\psi}(t)\}^T,-\int_{0}^C{G}_{K}(t,s)\hat{X}_i(s)\mathrm{d}s,\cdots, -\int_{0}^C{G}_{1}(t,s)\hat{X}_i(s) \mathrm{d}s)^T$,
$\bm{\nu}_i=(\bm{\alpha}_i^T,\bm{\omega}_{i}^T)^T$, 
and
$\bm{\zeta}_i(\bm{\beta})=(\sqrt{w(t_1)}\int_{0}^Cf_i(\bm{\hat{X}}(s),s;\bm{\beta})G_K(t_1,s) \mathrm{d}s,\cdots,\sqrt{w(t_H)}$ $\int_{0}^Cf_i(\bm{\hat{X}}(s),s;\bm{\beta})G_K(t_H,s)\allowbreak\mathrm{d}s)^T$.
In the following, we demonstrate procedures for solving \eqref{sim_lik}, which summary is given in Algorithm \ref{algo: GM}.

We first propose an optimization procedure for Green's matching under a separability condition.
Specially, we assume that $f_i(\bm{X}(t),t;\bm{\beta})$ satisfies
\begin{eqnarray}\label{de_f}
f_i(\bm{X}(t),t;\bm{\beta})=\bm{g}^T_i(\bm{X}(t),t)\ \bm{\beta}_i,\ i=1,\cdots,p,
\end{eqnarray}
where $\bm{g}_i(\cdot)$ is a known function and $(\bm{\beta}_1^T,\cdots,\bm{\beta}_p^T)^T=\bm{\beta}$.
This condition assumes that $f_i(\bm{X}(t),t;\bm{\beta})$ can be decomposed into a product of trajectory-related and parameter terms, a condition that is commonly adopted for statistical inference in dynamic systems \citep{brunton2016discovering,chen2017network, dai2021kernel}.
Under \eqref{de_f}, $\bm{\zeta}_i(\bm{\beta})$ in \eqref{sim_lik} satisfies
$\bm{\zeta}_i(\bm{\beta})=\bm{\Phi}_i\bm{\beta}_i$,
where $\bm{\Phi}_i=(\sqrt{w(t_1)}$ $\int_{0}^C\bm{g}_i(\bm{\hat{X}}(s),s)G_K(t_1,s) \mathrm{d}s,\cdots,\sqrt{w(t_H)} \int_{0}^C\bm{g}_i(\bm{\hat{X}}(s),s)G_K(t_H,s)\allowbreak\mathrm{d}s)^T$.
As a result, $\bm{\nu}_i$ and $\bm{\beta}_i$ can be estimated using the least squares fitting of \eqref{sim_lik} for each $i$.
This approach can be applied to cases such as the gene regulatory network, oscillatory dynamic direction model, and the harmonic equations in Section \ref{sim_dat_bo}, and also facilitates the equation discovery in Section \ref{ED}.

\begin{algorithm}
\caption{Green's matching for the case with unknown differential operators}
\label{algo: GM}
\begin{algorithmic}[1]
\State \textbf{Input} $t_j$, $j=1,\cdots,n$; $Y_{i}(t_j)$ $i=1,\cdots,p$ and $j=1,\cdots,n$. 
\State \textbf{For} $i=1,\cdots,p$
\State \quad Obtain $\hat{X}_i(t)$, $t=t_1,\cdots,t_H$, based on \eqref{local_est_1}.
\State \textbf{End for}
\State Set $\hat{\bm{X}}(t)=(\hat{X}_1(t),\cdots, \hat{X}_p(t))^T$, $t=t_1,\cdots,t_H$.
\State Calculate $\bm{Z}_i$, $\bm{\Psi}_i$, $i=1,\cdots,p$.
\State \textbf{If} the separability condition \eqref{de_f} is achieved,
\State \quad \textbf{For} $i=1,\cdots,p$ 
\State \quad\quad Calculate $\bm{\Phi}_i$ and set $\bm{R}_i=\ (\bm{\Psi}_i,\bm{\Phi}_i)$.
\State \quad\quad 
$(\bm{\nu}^T_i,\bm{\beta}^T_i)^T=\   (\bm{R}_i^T\bm{R}_i)^{-1}\bm{R}_i^T\bm{Z}_i$.
\State \quad \textbf{End for}
\State \quad $\bm{\beta}=\ (\bm{\beta}^T_1, \cdots, \bm{\beta}^T_p)^T$.
\State \textbf{Else} 
\State \quad Obtain $\bm{\beta}$ from the optimization \eqref{op_11}.
\State\quad $\bm{\nu}_i=(\bm{\Psi}_i^T\bm{\Psi}_i)^{-1}\bm{\Psi}_i^T\left\{\bm{Z}_i -\bm{\zeta}_i(\bm{\beta})\right\}.$
\State  \textbf{End if}
\State \textbf{Output} $\bm{\beta}$; $\nu_i$, $i=1,\cdots,p$.
\end{algorithmic}
\end{algorithm}

For general cases, the minimizer of \eqref{sim_lik} may not have a closed form, and we instead employ iteration methods to solve the optimization problem.
Letting $\bm{\nu}_i(\bm{\beta}):=(\bm{\Psi}_i^T\bm{\Psi}_i)^{-1}\bm{\Psi}_i^T\allowbreak\left\{\bm{Z}_i -\bm{\zeta}_i(\bm{\beta})\right\}$, the optimization \eqref{sim_lik} can be represented as
\begin{eqnarray}\label{op_11}
 \min_{\bm{\beta}\in\Omega_{\bm{\beta}}}\sum_{i=1}^p
\left\|\bm{Z}_i-\bm{\Psi}_i\bm{\nu}_i(\bm{\beta}) -\bm{\zeta}_i(\bm{\beta})\right\|^2,
\end{eqnarray}
and its derivative of $\bm{\beta}$ is
$-2\sum_{i=1}^p\big[\bm{\Psi}_i\left\{\nabla\bm{\nu}_i(\bm{\beta})\right\} +  \nabla\bm{\zeta}_i(\bm{\beta})\big]^T
\big\{\bm{Z}_i-\bm{\Psi}_i\bm{\nu}_i(\bm{\beta})-\bm{\zeta}_i(\bm{\beta})\big\}$,
where $\nabla$ is the gradient operator. 
Since $\nabla\bm{\nu}_i(\bm{\beta})=-(\bm{\Psi}_i^T\bm{\Psi}_i)^{-1}\bm{\Psi}_i^T\bm{\Phi}_i\left\{\nabla\bm{\zeta}_i(\bm{\beta})\right\}$, we can specify $\bm{\zeta}_i(\bm{\beta})$ and $\nabla\bm{\zeta}_i(\bm{\beta})$ to calculate the above gradient.
It's worth nothing that \eqref{op_11} can be considered a nonlinear least-squares problem. Therefore, it can be efficiently solved by many existing Newton-type methods given the above gradients; see \citet{nocedal1999numerical} for more details of these methods. 

\section{Statistical theory}\label{prope}
In this section, we investigate statistical properties of Green's matching and gradient matching for general-order dynamic systems, with the order $K$ to be fixed. 
For simplicity in notation, we only focus on the cases where $\mathcal{D}_i^k$s are known, and the bandwidths $h_k$s in (\ref{local_est_1}) are the same for different variables $i$. 
The results for cases with unknown differential operators and variable-specified bandwidths can be obtained similarly. 

Denote the true value of $\bm{\beta}$ as $\bm{\beta}_0$, and use $\bm{\hat{\beta}}^{\text{Grad}}_K$, $\bm{\hat{\beta}}^{\text{Grad}}_{K-1}$, and $\bm{\hat{\beta}}^{\text{Gree}}$ as the estimators for $\bm{\beta}$ based on \eqref{gra_mat}, \eqref{intm}, and \eqref{gre_11}, respectively.
We assume that the weight function $w(\cdot)$ for $\bm{\hat{\beta}}^{\text{Grad}}_K$, $\bm{\hat{\beta}}^{\text{Grad}}_{K-1}$, and $\bm{\hat{\beta}}^{\text{Gree}}$ has compact support on $(0,C)$ and is positive and bounded on its support.

\subsection{Statistical properties of Green's matching}
Recall that $\{Y_{i}(t_j);i=1,\cdots,p, j=1, \cdots,n\}$ are generated from model (\ref{mea_model}), where $X_i(\cdot)$s satisfy \eqref{dyn} with $\bm{\beta}=\bm{\beta}_0$ and $\epsilon_{i}(t_j)$ is the while noise.
To obtain the consistency of Green's matching, we impose additional assumptions as follows:

\begin{Aass}\label{A.1}
The variances of $\varepsilon_{i}(t)$s, denoted as $\sigma^2_i(t)$s, are continuous function of $t$ on $[0,C]$.
\end{Aass}

\begin{Aass}\label{A.2}
 The observed time points $\{t_{j}; j=1,\cdots, n\}$ are the i.i.d. random variables with density function $f_T(t)$, where $f_T(t)$ is bounded away from $0$ and continuous on $[0,C]$.
\end{Aass}

\begin{Aass}\label{A.3}
The weight function $W(\cdot)$ used in in Section \ref{pres} is symmetric about the origin and has a compact support.
\end{Aass}

\begin{Aass}\label{A.4}
For each $i$, $X_i(t)$ has a uniformly bounded second-order derivatives on $[0,C]$. 
\end{Aass}

Assumptions \ref{A.1}, \ref{A.2}, and \ref{A.3} are commonly assumed in the literature to establish the consistency of local polynomial regressions \citep{stone1982optimal,fan2018local}. Additionally, we impose Assumption \ref{A.4} to ensure the uniform convergence of the local polynomial fitting \citep{stone1982optimal}.

Without loss of generality, we assume that for each $i$, the driving function $f_i(\bm{X}(t),t;\bm{\beta})$ is time-invariant, i.e., $f_i(\bm{X}(t),t;\bm{\beta})= f_i(\bm{X}(t);\bm{\beta})$, $\forall t\in [0,C]$.
For the time-varying case, we can replace $\bm{X}(t)$ with $(\bm{X}(t),t)^T$ in $f_i(\bm{X}(t);\bm{\beta})$.

\begin{Bass}\label{B.1}
$f_i(\bm{X}(t) ;\boldsymbol{\beta})=f_i(\bm{X}(t)  ;\boldsymbol{\beta}^{\prime})$ for all $i$ and $t$ if and only if $\boldsymbol{\beta}=\boldsymbol{\beta}^{\prime}$.
\end{Bass}

\begin{Bass}\label{B.2}
The parameter space $\Omega_{\bm{\beta}}$ is a compact subset of a Euclidean space, and the initial values $\{D^k{X}_i(0);k\leq K,i\leq p\}$ are uniformly bounded.
\end{Bass}

\begin{Bass}\label{B.3}
For each $i$,
${f}_i(\cdot;\cdot)$ is continuous on $\mathbb{R}^{p}\times \Omega_{\boldsymbol{\beta}}$.
\end{Bass}

Assumption \ref{B.1} introduces identifiability for the parameter estimation. 
Furthermore, according to \eqref{green_inverse_2}, Assumption \ref{B.2} is equivalent to having compact parameter spaces $\Omega_{\bm{\alpha}}$ and $\Omega_{\bm{\beta}}$ in \eqref{gre_11}. Finally, Assumption \ref{B.3} introduces smoothness for $f_i(\cdot;\cdot)$. In the literature, stronger conditions, such as the uniformly Lipschitz condition or the continuously differentiable condition \citep{brunel2008parameter,Shota}, have been imposed on $f_i(\cdot;\cdot)$ to control estimation errors. We alleviate these conditions by assuming only a continuity condition, taking advantage of the uniform convergence of local polynomial fitting under Assumption \ref{A.4}.

\begin{theorem}[Consistency]
Under Assumptions \ref{A.1}-\ref{A.4} and \ref{B.1}-\ref{B.3}, and the bandwidth $h_0$ satisfying
$h_0=o(1)$ and  $n^{-1}\ln n/h_0\to 0$ as $n \to \infty$, then
\begin{eqnarray*}
    \left\|\boldsymbol{\hat{\beta}}^{\text{Gree}}-\boldsymbol{\beta}_0\right\|=o_P(1),
\end{eqnarray*}
as $n \to \infty$, i.e. $\boldsymbol{\hat{\beta}}^{\text{Gree}}$ converges to $\boldsymbol{\beta}_0$ in probability. \label{thm1}
\end{theorem}
To further obtain the asymptotic normality of $\boldsymbol{\hat{\beta}}^{\text{Gree}}$, we additionally assume that: 

\begin{Bass}\label{B.4}
For each $i$, $\frac{\partial f_i(\boldsymbol{X};\boldsymbol{\beta})}{\partial \boldsymbol{X}}, \frac{\partial f_i(\boldsymbol{X};\boldsymbol{\beta})}{\partial \boldsymbol{\beta}}, \frac{\partial^{2} f_i(\boldsymbol{X};\boldsymbol{\beta})}{\partial \boldsymbol{X}\partial \boldsymbol{X}^{T}},
 \frac{\partial^{2} f_i(\boldsymbol{X};\boldsymbol{\beta})}{\partial \boldsymbol{\beta} \partial \boldsymbol{\beta}^{T}}$
 are continuous on  $\mathbb{R}^{p}\times \Omega_{\boldsymbol{\beta}}$, where we abuse the notation $\boldsymbol{X}$ to denote the first variable of $f_i(\cdot; \cdot)$.
\end{Bass}

\begin{Bass}\label{B.5}
The matrix $\boldsymbol{\Sigma}_1=\frac{\partial^{2} \mathcal{L}^{\text{Gree}}(\bm{\theta}_0)}{\partial \boldsymbol{\theta} \partial \boldsymbol{\theta}^{T}}$
is positive definite, where $\boldsymbol{\theta}=\left(\boldsymbol{\beta}^T,\boldsymbol{\alpha}^T \right)^T$ with $\boldsymbol{\theta}_0$ being its true value, and
$
\mathcal{L}^{\text{Gree}}(\boldsymbol{\theta})=\int_{0}^C\sum_{i=1}^p\left[\mathcal{T}^{i,\text{Gree}}_{t,\bm{\beta},\bm{\alpha}_i}(\bm{{X}})\right]^2w(t)\ \mathrm{d}t
$.
\end{Bass}

Assumption \ref{B.4} imposes additional smoothness conditions to bound the error terms of $\hat{\bm{\beta}}^{\text{Gree}}$, which is a conventional condition as assumed in \citet{brunel2008parameter}. 
Moreover, Assumption \ref{B.5} is called the local identifiability condition \citep{brunel2008parameter}, as $\bm{\Sigma}_1$ is approximately the Hessian matrix of the likelihood function in \eqref{gre_11} at the true parameter. Under this condition, the likelihood function is well-separated around the true parameter since it has a strictly positive curvature at $\boldsymbol{\theta}_0$.

In the following, we denote the $u^{th}$ component of $\frac{\partial f_i(\boldsymbol{X};\boldsymbol{\beta})}{\partial \boldsymbol{X}}$ as $\frac{\partial f_i(\boldsymbol{X};\boldsymbol{\beta})}{\partial X_{u}}$. The notations $\frac{\partial f_{i}(\boldsymbol{X}(s);\boldsymbol{\beta}_0)}{\partial \boldsymbol{\beta}}$ and $\frac{\partial f_{i}(\boldsymbol{X}(s);\boldsymbol{\beta}_0)}{\partial X_u}$ are also simplified as $\frac{\partial f_{i}(s)}{\partial \boldsymbol{\beta}}$ and $\frac{\partial f_{i}(s)}{\partial X_u}$, respectively. Besides, we define
$\boldsymbol{d}_{i}(t)=(\int_{0}^1 G_i^K(t,s)\frac{\partial f_{i}(s)}{\partial \boldsymbol{\beta}^T}\mathrm{d}s,\ \boldsymbol{e}_i^T \otimes  \{\bm{\psi}_i(t)\}^T )^T$, where $\boldsymbol{e}_i\in \mathbb{R}^p$ satisfies $[\boldsymbol{e}_i]_{i^{\prime}}=\mathbb{I}(i=i^{\prime})$ and $\otimes$ is the Kronecker product.

\begin{theorem}[Asymptotic normality]
Under Assumptions \ref{A.1}-\ref{A.4}, \ref{B.1}-\ref{B.5}, we assume that the bandwidth $h_0$ satisfies
$
h_0=o(n^{-\frac{1}{4}})$ and $n^{-\frac{1}{2}}/h_0\to 0$ as $n\rightarrow \infty$. We then establish the asymptotic normality of $\boldsymbol{\hat{\beta}}^{\text{Gree}}$, i.e.,
\begin{eqnarray*}
\sqrt{n}\left(\boldsymbol{\hat{\beta}}^{\text{Gree}}-\boldsymbol{\beta}_0\right) \to _{d} \mathcal{N}(\boldsymbol{0},\boldsymbol{\Sigma}),
\end{eqnarray*}
where $\boldsymbol{\Sigma}$ is the first $m\times m$ sub-matrix of $\boldsymbol{\Sigma}_1^{-1} \boldsymbol{\Sigma}_2 \boldsymbol{\Sigma}_1^{-1}$ with $m$ being the number of the parameters $\bm{\beta}$, and
\begin{eqnarray*}
    \bm{\Sigma}_{2}=4\sum_{i,l,u=1}^{p}     \int_0^1\int_0^1\int_0^1 \frac{\sigma^2_i(\tau)}{f_T(\tau)}
    \boldsymbol{H}_{i,l}(t,\tau)\cdot\left\{\boldsymbol{H}_{i,u}(s,\tau)\right\}^T \ \mathrm{d}t \mathrm{d}s \mathrm{d}\tau\nonumber,
\end{eqnarray*}
among $\bm{H}_{i,l}(t,\tau)=G^K_l(t,\tau)\frac{\partial f_{l}(\tau)}{\partial X_i}\bm{d}_l(t)w(t)-\frac{1}{p}\bm{d}_i(\tau)w(\tau)$.

\label{thm2}
\end{theorem}

We immediately observe that bandwidths such as $h_0=n^{\alpha}\ln^c n$ (with $\alpha\in (-1/2,-1/4)$ and $c \in \mathbb{R}$), $h_0=n^{-\frac{1}{2}}\ln^{c} n$ (for $c >0$), and $h_0=n^{-\frac{1}{4}}\ln^{c} n$ (for $c <0$) satisfy the bandwidth assumption in Theorem \ref{thm2}. It is worth noting that these bandwidths are smaller than the typical order $n^{-\frac{1}{5}}$ in the regular local polynomial fitting, leading to the undersmoothing on the curves. The undersmoothing requirement is common in two-step methods to decrease the estimation bias raised by the first-step estimation \citep{carroll1997generalized,zhou2019efficient}. 
Moreover, a direct consequence of Theorem \ref{thm2} is that
\begin{eqnarray*}
    \|\boldsymbol{\hat{\beta}}^{\text{Gree}}-\boldsymbol{\beta}_0\|=O_P(n^{-\frac{1}{2}}),
\end{eqnarray*}
where the root-$n$ consistency of $\boldsymbol{\hat{\beta}}^{\text{Gree}}$ is independent of the differential-order $K$, indicating that Green's matching is optimal in parameter estimations for general-order dynamic systems. This superior theoretical property comes from integrating Green's functions in our estimation procedure, which transforms the differential equation into the integral equation with zero-differential orders. As a result, Green's matching avoids estimating higher-order derivatives that tend to possess significant biases \citep{fan2018local}, achieving the goal of de-biasing in the estimation process.

\subsection{Convergence rates of three methods}\label{bias_rat}
In this subsection, we compare the estimation biases and convergence rates of ${\bm{\hat{\beta}}}^{\text{Gree}}$, ${\bm{\hat{\beta}}}^{\text{Grad}}_K$, and ${\bm{\hat{\beta}}}^{\text{Grad}}_{K-1}$ under the general-order system framework.
Since the gradient matching for high-order systems requires estimating derivatives of $X_i(\cdot)$, we impose additional assumptions to control the pre-smoothing errors.
\begin{Aass}\label{A.6}
For each $i$, $X_i(\cdot)$ has a uniformly bounded $(K+2)^{th}$ derivative on $[0,C]$.
\end{Aass}

\begin{Bass}\label{B.6}
The matrix $\boldsymbol{\Sigma}'_1=\frac{\partial^{2} \mathcal{L}^{\text{Grad}}_K(\boldsymbol{\beta}_0)}{\partial \boldsymbol{\beta} \partial \boldsymbol{\beta}^{T}}$
is positive definite, where 
$$
\mathcal{L}^{\text{Grad}}_K(\boldsymbol{\beta})=\sum_{i=1}^p\int_{0}^C\bigg\{
\mathcal{T}^{i,K}_{t,\bm{\beta},\bm{\omega}}(\bm{X}, D\bm{X}, \cdots,D^K\bm{X})\bigg\}^2 w(t)\ \mathrm{d}t.
$$
\end{Bass}

\begin{Bass}\label{B.7}
The matrix $\boldsymbol{\Sigma}''_1=\frac{\partial^{2} \mathcal{L}^{\text{Grad}}_{K-1}(\bm{\theta}_0)}{\partial \boldsymbol{\theta} \partial \boldsymbol{\theta}^{T}}$
is positive definite, where $$
\mathcal{L}^{\text{Grad}}_{K-1}(\boldsymbol{\theta})=\sum_{i=1}^p\int_{0}^C\bigg\{
\mathcal{T}^{i,\text{Grad},K-1}_{t,\bm{\beta},\bm{\omega},\alpha_i}(\bm{X}, D\bm{X}, \cdots,D^{K-1}\bm{X})\bigg\}^2w(t)\ \mathrm{d}t,$$ and we abuse $\boldsymbol{\theta}$ to denote 
$\boldsymbol{\theta}=\left(\boldsymbol{\beta}^T,\alpha_1,\cdots,\alpha_p \right)^T$ with $\boldsymbol{\theta}_0$ being its true value.

\end{Bass} 

Define the estimation bias for an estimator $\bm{\hat{\beta}}$ as $\textbf{Bias}({\bm{\hat{\beta}}} ):=\|\mathbb{E}{\bm{\hat{\beta}}}-\bm{\beta}_0\|$.

\begin{theorem}[Estimation biases and rates of convergence] 
$\ $

(a) Under Assumptions \ref{A.1}-\ref{A.4}, \ref{B.1}-\ref{B.5}, we assume that $h_0$ satisfies $h_0=o(1)$ such that  $n^{-1}\ln n/h_0\to 0$ as $n \to \infty$, then 
\begin{eqnarray*}
    \textbf{Bias}(\bm{\hat{\beta}}^{\text{Gree}})=O\bigg(h_{0}^2+n^{-1}h_0^{-1}\bigg)\ \text{and}\ \|\bm{\hat{\beta}}^{\text{Gree}}-\bm{\beta}_0\|=O_P\left(h_{0}^2+n^{-1}h_0^{-1}+n^{-\frac{1}{2}}\right).
\end{eqnarray*}

(b) Under Assumptions \ref{A.1}-\ref{A.3}, \ref{A.6}, \ref{B.1}-\ref{B.4}, \ref{B.6}, \ref{B.7}, and $h_{k}=o(1)$ such that $n^{-1}\ln n/ h_k^{2{k}+1}\to 0$ as $n \to \infty$ for ${k}\leq K_0$, where $K_0=K$ or $K-1$, then
\begin{eqnarray*}
    \sqrt{n}\left( \bm{\hat{\beta}}^{\text{Grad}}_{K_0}-\bm{\beta}_0\right)=\bm{Z}_{n,K_0}+\bm{W}_{n,K_0},
\end{eqnarray*}
where $\bm{Z}_{n,K_0}$ converges to a centered Gaussian vector in distribution, and
$$\bm{W}_{n,K_0}=O_P\bigg\{\sum_{k=0}^{K_0}(n^{\frac{1}{2}}h_{k}^2+n^{-\frac{1}{2}}h_0^{-\frac{1}{2}}h_{k}^{-k-\frac{1}{2}})\bigg\}$$ involves the error terms relating to the pre-smoothed curves $\bm{\hat{X}}, \widehat{D\bm{X}},  \cdots,\widehat{D^{K_0}\bm{X}}$. As a result, we have
\begin{eqnarray*}
    &\textbf{Bias}(\bm{\hat{\beta}}^{\text{Grad}}_{K_0})= O\bigg\{\sum_{k=0}^{K_0}(h_{k}^2+n^{-1}h_0^{-\frac{1}{2}}h_{k}^{-k-\frac{1}{2}})\bigg\},\\ &\|\bm{\hat{\beta}}^{\text{Grad}}_{K_0}-\bm{\beta}_0\|=O_P\bigg\{\sum_{k=0}^{K_0}(h_{k}^2+n^{-1}h_0^{-\frac{1}{2}}h_{k}^{-k-\frac{1}{2}})+n^{-\frac{1}{2}}\bigg\}
\end{eqnarray*}
for $K_0=K$ or $K-1$. 
\label{thm3}
\end{theorem}

Combining with Theorems \ref{thm2} and \ref{thm3}, the main difference of the asymptotic behaviors between Green's matching and gradient matching lies in the presence of  $\bm{W}_{n,K_0}$, where the  convergence rate gets slow as $K_0$ increases. Based on this, the convergence orders of the estimation biases increase in the order of Green's matching, order $(K-1)$ gradient matching, and order $K$ gradient matching, where the corresponding used pre-smoothing is increased in turn. 
Typically, the pre-smoothing step would enlarge the total bias of the resulting estimator by ignoring prior information of the underlying model. Therefore, it can be expected that the larger amount of pre-smoothing used in the first step, the greater bias in the second-step estimation.

Furthermore, when choosing optimal bandwidths in Theorem \ref{thm3}, we can further specify the convergence rates of the gradient matching with different orders. Specifically, when $h_0,\ h_1, \cdots,\ h_{K}$ are of the order $n^{-\frac{1}{K+3}}$, the order $K$ gradient matching achieves the lower bound
\begin{eqnarray*}
    \|\bm{\hat{\beta}}^{\text{Grad}}_K-\bm{\beta}_0\|=O_P\left(n^{-\frac{2}{K+3}}+n^{-\frac{1}{2}}\right).
\end{eqnarray*}
Similarly, the convergence rate of $\bm{\hat{\beta}}^{\text{Grad}}_{K-1}$ attains its lower bound
\begin{eqnarray*}
    \|\bm{\hat{\beta}}^{\text{Grad}}_{K-1}-\bm{\beta}_0\|=O_P\left(n^{-\frac{2}{K+2}}+n^{-\frac{1}{2}}\right)
\end{eqnarray*}
when $h_0,\ h_1, \cdots,\ h_{K-1}$ are of the order $n^{-\frac{1}{K+2}}$.
If $K=1$, these results are consistent with \citet{brunel2008parameter}, \citet{Shota}, and \citet{Itai} for first-order systems. However, as $K$ becomes larger, the biases would dominate the convergence rates, leading to the order $K$ or $(K-1)$ gradient matching may not achieve $\sqrt{n}$-consistency when $K \geq 2$ or $K \geq 3$, respectively.
Meanwhile, Green's matching can always obtain the $\sqrt{n}$-consistency.

\section{Simulation}\label{sim_dat_bo}
\subsection{Comparisons of two-step methods}\label{sim_1}
In this subsection, we examine the estimation behaviors of three two-step methods for parameter estimation in general-order dynamic systems. For the comparison, we implement the two-step methods with the same pre-smoothing using local polynomial regressions, and we select the bandwidths through a data-driven cross-validation method \citep{fan2018local}; refer to Part 1.1 in Supplementary materials for the detailed implementation.
Furthermore, we use three test models to compare the performances of different methods, including one first-order and two second-order dynamic systems. For these cases, we only focus on systems with a relatively large $p$, for which the two-step methods are usually employed for parameter estimation.
The details of these dynamic systems are listed below. In all cases, we focus on the functions on $[0, C]$ with $C$ set to $20$.

{\textbf{Gene regulatory network.}}
A gene regulatory network is a collection of molecular regulators that interact with each other to determine the function of a cell. We consider dynamic systems with an additive structure to model a gene regulatory network \citep{wu2014sparse}, where the state variables are assumed to follow
\begin{eqnarray*}
DX_{i}(t)= a_{i1}g_{i1}(X_{i}(t))+a_{i2}g_{i2}(X_{i+5}(t))+a_{i3}g_{i3}(X_{i-3}(t)),
\end{eqnarray*}
among $X_i(t)$ is the concentration of the $i^{th}$ gene, and we suppose that $X_i(t)=X_{i-p}(t)$ when $i>p$ and $X_i(t)=X_{i+p}(t)$ when $i<1$. We set that $p=50$, $g_{i1}(x)=\sin(\frac{1}{2}x)$,
$g_{i2}(x)=-\cos(\frac{1}{2}x)$, and $g_{i3}(x)=x$. Besides, we assume $a_{i1}=0.5$, $a_{i2}=1$, $a_{i3}=0.1$, and $X_i(0)=5 -\frac{4}{49}\cdot (i-1)$ for $i=1,\cdots,p$.
This model contains 150 parameters to be estimated from data, including $a_{i1}$, $a_{i2}$, and $a_{i3}$ for $i=1,\cdots,50$.

{\textbf{Spring-mass system.}}
We consider a second-order dynamic system constructed by Newton's second law: the spring-mass system. The spring-mass system is a Newtonian mechanical model that describes the motion of a system connected by springs. We consider $p$ objects with unit masses. According to Newton's second law \citep{spring-mass}, the motion equations can be represented as follows:
\begin{eqnarray*}
D^2X_1(t)&=&c_2\left(X_2(t)-X_1(t)-L_2
\right) - c_1\left(X_1(t)-L_1\right) + g,\\
D^2X_i(t)&=&c_{i+1}(X_{i+1}(t)-X_i(t)-L_{i+1})
-  c_{i}(X_{i}(t)-X_{i-1}(t) -L_i)+g,\\
& &i=2,\cdots,p-1,\\
D^2X_p(t)&=&-c_{p}\left(X_{p}(t)-X_{p-1}(t)-L_i\right)+ g + 5\sin(t),
\end{eqnarray*}
where $X_i(t)$ is the position of the $i^{th}$ object at time $t$ in a one-dimensional coordinate system, $g$ is the gravitational acceleration, $c_i$ is the spring constant of the $i^{th}$ spring, $L_i$ is the undeformed length of the $i^{th}$ spring, and $5\sin(t)$ is a time-varying forcing applied to the $p^{th}$ object. We set that $L_i=2$, $c_i=3$, $i=1,\cdots, p$ with $p=10$, $g=9.8$, $X_i(0)=2i$, and $DX(0)=0$. Assuming $L_i$, $i\leq p$, and $g$ are known, we want to estimate the spring constants $c_i$.

{\textbf{Oscillatory dynamic directional model.}}
Dynamic directional models are the systems that characterize interactions among neuronal states. Here, we consider an oscillatory dynamic directional model with damped harmonic oscillators to depict the oscillatory activity of neuronal systems \citep{zhang2020bayesian}. The model is 
\begin{eqnarray*}
D^2X_i(t)+a_{i}DX_i(t)+b_{i}X_i(t)
&=& c_{i}X_{i+1}(t)+ d_{i} \mathbb{I}(2\leq t \leq 3),\ \text{for}\ i=1,\cdots, p-1, \\
D^2X_p(t)+a_{p}DX_p(t)+b_{p}X_p(t)&=&
 d_{p} \mathbb{I}(2\leq t \leq 3),
\end{eqnarray*}
where $X_i(t)$ is the $i^{th}$ neuronal state functions at time $t$, $a_{i}$ and $b_{i}$ are the positive parameters that determine the shape of the $i^{th}$ curve, $c_{i}$ is the rate of the directed effect from the $(i+1)^{th}$ neuronal state onto the $i^{th}$ neuronal state, and $d_{i} \mathbb{I}(2\leq t \leq 3)$ is the time-varying signal that activates the $i^{th}$ neuronal state. We assume that $p=50$, and the parameters $a_{i}$, $b_{i}$, $c_{i}$, and $d_{i}$, in total 199 parameters, are estimated from data. We set that $a_{i}=0.4+{0.2\cdot (i-1)}/ {49}$, $b_{i}=0.9+{0.2\cdot (i-1)}/ {49}$, $c_{i}=0.2\cdot (-1)^{i+1}$, $d_{i}=0.1$, and $(X_i(0),DX_i(0))=(0,0)$ for different $i$. 

For the above equations, we simulate their dynamic curves using the corresponding true values of parameters and the initial conditions mentioned above. The Figure 1 in Supplementary Materials shows the simulated dynamic curves.
Next, we generate $Y_{i}(t_j)$ from a normal distribution with mean $X_i(t_j)$ and variance $\sigma_i^2$, where $\sigma_i^2=\gamma^2\int_0^CX_i^2(t)\ \mathrm{d}t /C$ with $\gamma$ controlling the contamination level.
It's worth noting that \cite{wu2014sparse} and \cite{zhang2020bayesian} inferred the gene regulatory network and the oscillatory dynamic directional model by gradient matching. Here, we further examine the estimation behaviors of Green's matching for these dynamic systems.

\begin{table}[ht!]
\centering
\renewcommand{\arraystretch}{1.1}
\caption{Comparison of RRMSEs of three methods.}
\begin{tabular}{l|c|c|c|c}
   \hline
\multicolumn{2}{c|}{ {RRMSE(\%)}}
   &  \multicolumn{1}{c|}{$\gamma=3\%$} &  \multicolumn{1}{c|}{$\gamma=5\%$}& \multicolumn{1}{c}{$\gamma=7\%$}\\
   \hline
  \multicolumn{5}{c}{Gene regulatory network}\\
  \hline
  \multirow{2}{*}{$n=50$} 
     & Order 1 gradient matching & 19.04 & 25.19 & 30.30 \\ 
     & Green’s matching & 9.74 & 15.23 & 20.34 \\ 
             \hline
               \multirow{2}{*}{$n=150$} 
         & Order 1 gradient matching & 14.15 & 18.75 & 22.74 \\ 
       & Green’s matching & 5.95 & 9.33 & 12.54 \\
               \hline
                 \multirow{2}{*}{$n=250$} 
      & Order 1 gradient matching & 12.40 & 16.24 & 19.56 \\ 
         & Green’s matching & 4.78 & 7.51 & 10.12 \\ 
         \hline
         \multicolumn{5}{c}{Spring-mass system}\\
  \hline
       \multirow{3}{*}{$n=50$}      
 & Order 2 gradient matching & 61.36 & 72.71 & 78.54 \\ 
     & Order 1 gradient matching & 18.62 & 32.67 & 43.71 \\ 
      & Green’s matching & 13.18 & 21.67 & 29.57 \\ 
      \hline
          \multirow{3}{*}{$n=150$}  
   & Order 2 gradient matching & 44.05 & 59.48 & 67.84 \\ 
        & Order 1 gradient matching & 8.97 & 17.37 & 25.41 \\ 
         & Green’s matching & 6.19 & 11.86 & 16.44 \\ 
 \hline
           \multirow{3}{*}{$n=250$}  
 & Order 2 gradient matching & 36.02 & 51.57 & 61.15 \\ 
           & Order 1 gradient matching & 6.50 & 12.50 & 19.21 \\ 
            & Green’s matching & 4.57 & 8.29 & 13.66 \\ 
         \hline    
    \multicolumn{5}{c}{Oscillatory dynamic directional model}\\
         \hline
      \multirow{3}{*}{$n=50$}      
& Order 2 gradient matching & 70.57 & 87.73 & 104.90 \\ 
     & Order 1 gradient matching & 37.78 & 49.31 & 60.43 \\ 
      & Green’s matching & 9.64 & 15.90 & 22.29 \\ 
      \hline
          \multirow{3}{*}{$n=150$}  
    & Order 2 gradient matching & 74.60 & 86.26 & 96.41 \\ 
        & Order 1 gradient matching & 29.12 & 37.81 & 45.10 \\ 
         & Green’s matching & 5.50 & 9.11 & 12.99 \\ 
 \hline
           \multirow{3}{*}{$n=250$}  
    & Order 2 gradient matching & 70.23 & 80.98 & 89.55 \\ 
           & Order 1 gradient matching & 25.76 & 33.07 & 39.26 \\ 
            & Green’s matching & 4.33 & 7.08 & 10.02 \\
            \hline
\end{tabular}\label{ODE1_result}
\end{table} 

We use the order $K$ and $(K-1)$ gradient matching and Green's matching to fit the observed data. 
Since Green's matching is equivalent to order $(K-1)$ gradient matching in the gene regulatory network, we only compare the estimation behaviors of Green's matching and order $K$ gradient matching in this case. 

We define a measure to compare the estimation errors of different methods. Let $\beta_{l}$ be the true value of the $l^{th}$ parameter (including those in the driving functions and differential operators), and ${\beta}^{(g)}_l$ be the corresponding estimators in the $g^{th}$ simulation, where $l\leq m$ and $g\leq G$ with $m$ and $G$ being the number of parameters and the replicated simulations, respectively. We define the root relative mean square error (RRMSE) as follows
\begin{eqnarray*}
    \text{RRMSE}:=\frac{1}{m}\sum_{l=1}^m\frac{\sqrt{\frac{1}{G}\sum_{g=1}^G(\beta_l^{(g)}-{\beta}_l)^2}}{\beta_l}\cdot 100\%
\end{eqnarray*}
The results with $G=100$ and different values of $n$s and $\gamma$s are given in Table \ref{ODE1_result}. 
We find that Green's matching outperforms the gradient matching at different orders for all the test models under different settings.
In particular, the gradient matching approaches exhibit poor performances for second-order systems in cases with a high contamination level $\gamma$, where the pre-smoothed derivatives are usually inaccurate. 
In contrast, Green's matching maintains satisfactory performances in such scenarios.
These results suggest that the pre-smoothed derivatives significantly impact the estimation accuracy of the two-step procedures. In this regard, Green's matching removes all the pre-smoothed derivatives for the second-step estimation, thereby obtaining a better estimation performance than the other two methods.

\begin{figure}[h]
\centering
\includegraphics[scale = 0.4]{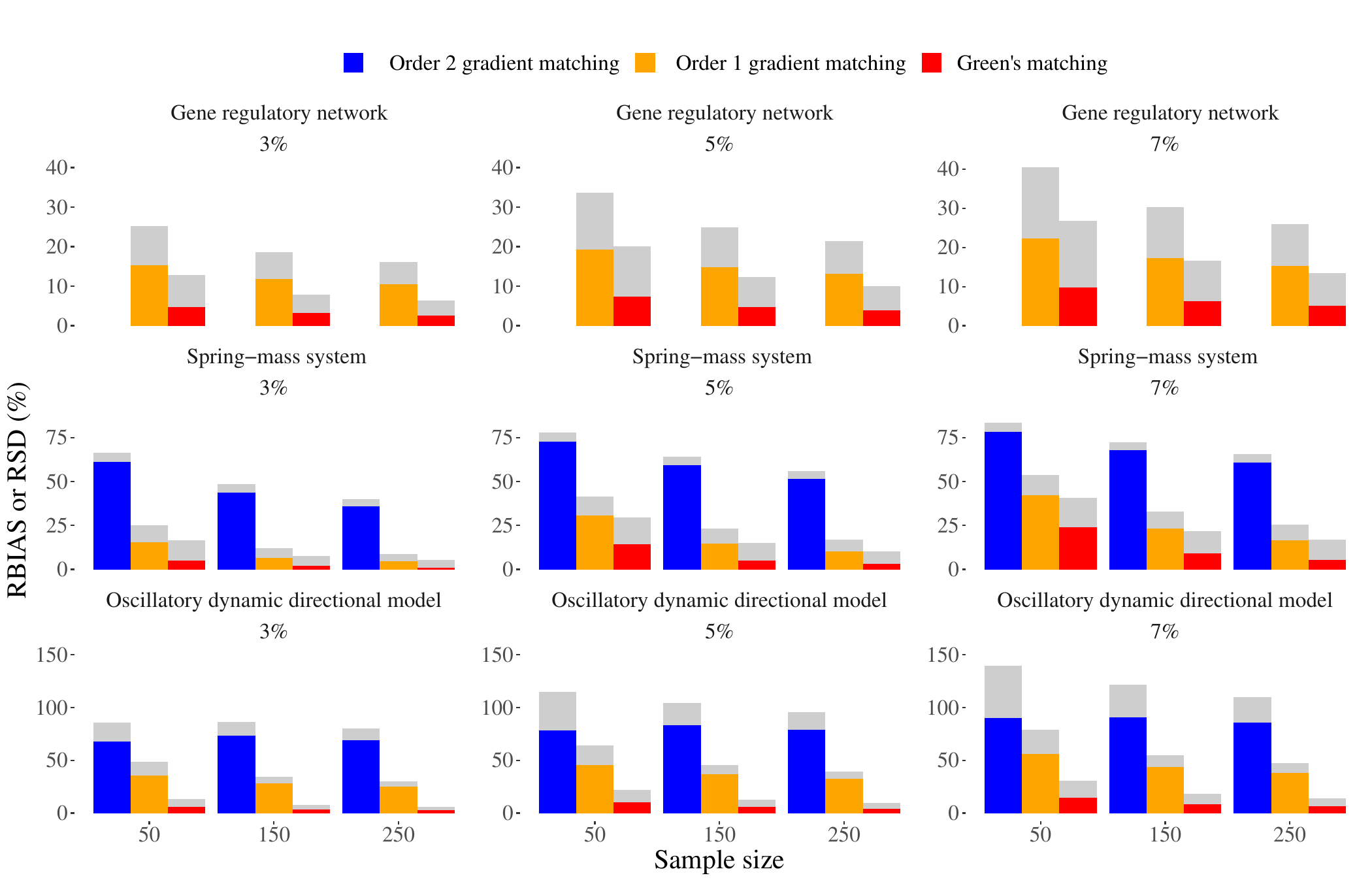}
\caption{RBIAS and RSD of three methods with different sample sizes $n$ and $\gamma$s (sub-title), where RSDs are colored in gray, and RBIASs of order 2 / 1 gradient matching and Green's matching are colored in blue, orange, and red, respectively.}\label{para_est}
\end{figure}

To further evaluate the estimation performance, we separate RRMSE into two quantities, which are 
\begin{eqnarray*}
\text{RBIAS}:=\frac{1}{m}\sum_{l=1}^m\left|\frac{\beta_l-\bar{\beta}_l}{\beta_l}\right|\cdot 100\%\ \text{and}\ 
\text{RSD}:=\frac{1}{m}\sum_{l=1}^m\frac{\sqrt{\frac{1}{G}\sum_{g=1}^G(\beta_l^{(g)}-\bar{\beta}_l)^2}}{\beta_l}\cdot 100\%,
\end{eqnarray*}
where $\bar{\beta}_l=\frac{1}{G}\sum_{g=1}^G{\beta}^{(g)}_l$. RBIAS and RSD are two additional measures to assess the corresponding estimators' relative estimation biases and variances. In Figure \ref{para_est}, we present the RBIASs and RSDs of the three methods for all models. We find that the estimation biases of gradient matching always dominate their estimation errors, whereas those of Green's matching are nearly balanced with the corresponding RSDs. These results are consistent with Theorem \ref{thm3}. For a two-step estimator, the pre-smoothing process tends to enlarge the total estimation bias, leading to a poor convergence and/or finite-sample behavior for parameter estimation. However, these estimation biases can be alleviated by using Green's matching, which results in superior performances for parameter estimation in general-order systems.

\begin{figure}[h]
\centering
\includegraphics[scale = 0.5]{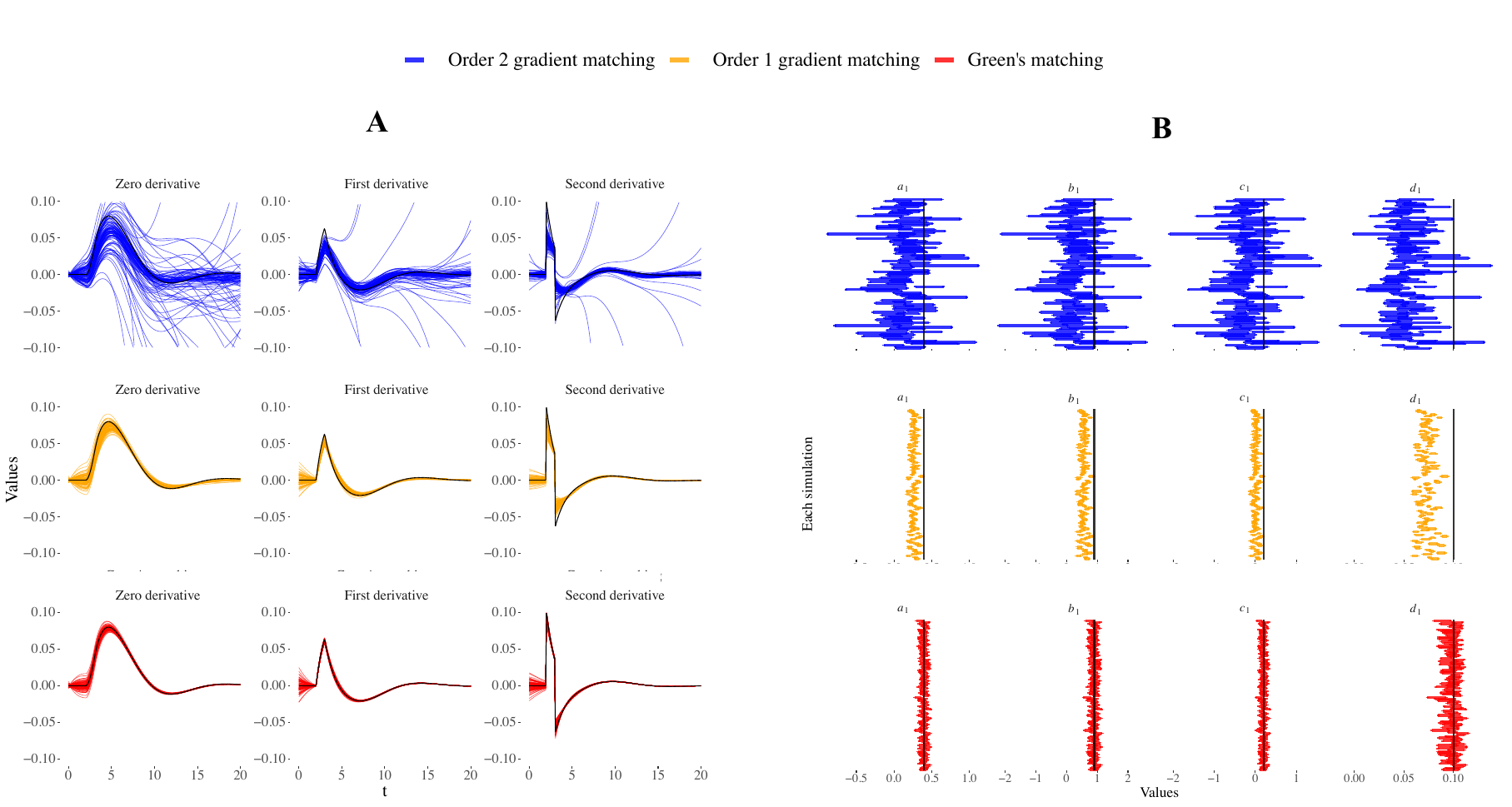}
\caption{\textbf{A}. Reconstructed curves for $X_1(\cdot)$, $DX_1(\cdot)$, and $D^2X_1(\cdot)$ (in the oscillatory dynamic directional model) by three methods from 100 simulations, where the solid black lines indicate the true curves.
\textbf{B}. Confidence intervals of $a_{1}$, $b_{1}$, $c_{1}$, and $d_{1}$ (in the oscillatory dynamic directional model) estimated by three methods from 100 simulations, where the solid black lines indicate the true values of parameters.}\label{para_est_1}
\end{figure}

For subsequent analysis, we only focus on the oscillatory dynamic directional model with $n=250$ and $\gamma=7\%$. To illustrate the ability to capture data patterns, we use a numerical approximation to reconstruct the trajectories $X_1(\cdot)$, $DX_1(\cdot)$, and $D^2X_1(\cdot)$ based on the pre-smoothed initial conditions and the estimated parameters obtained from the three methods. The results are shown in Figure \ref{para_est_1} \textbf{A}. For these cases, the order 2 gradient matching introduces $\widehat{D^2X}_1(\cdot)$ to the second-step parameter estimation, even though $D^2X_1(t)$ is discontinuous at $t=2$ or $t=3$. We observe that the reconstructed curves by order 2 gradient matching poorly fit their true shapes. In contrast, those by the other two methods appropriately fit the underlying curves, as they remove the pre-smoothed $D^2X_1(t)$ from the parameter estimation. These results suggest that incorporating pre-smoothing of non-smooth derivatives into the two-step methods may result in significant errors in parameter estimation.

Furthermore, we calculate the 95\% confidence intervals for the parameters $a_{1}$, $b_{1}$, $c_{1}$, and $d_{1}$ in the oscillatory dynamic directional model, utilizing the asymptotic normality of Green's matching in Theorem \ref{thm2}. We also employ a similar strategy to construct the 95\% confidence intervals by gradient matching of different orders. These 95\% confidence intervals are shown in Figure \ref{para_est_1}. \textbf{B}. In our simulation, nearly 95\% (94\%, 92\%, 96\%, 88\% for $a_{1}$, $b_{1}$, $c_{1}$, and $d_{1}$, respectively) of the confidence intervals constructed by Green's matching cover the true parameters. While the confidence intervals constructed by gradient matching are significantly biased with incorrect covering proportions (correspondingly, 39\%, 42\%, 47\%, and 7\% in order $K$ gradient matching, and 0\%, 0\%, 1\%, and 0\% in order $(K-1)$ gradient matching). 

In Parts 3.3 and 3.4 of Supplementary Materials, we provide the details of reconstructing trajectories and calculating confidence intervals as mentioned above. Additionally, we include the cases of $n=250$ and $\gamma=3\%$ or $5\%$ for the above model to facilitate more comparisons. These results further illustrate the competitive performance of Green's matching. Overall, Green's matching not only obtains a satisfactory estimation for the parameters to capture data patterns but also provides well-behaved confidence intervals that appropriately cover the true parameters.

\subsection{Comparisons of other competitive methods}\label{comp_sim_2}
In this subsection, we propose further comparisons with two methods that do not involve pre-smoothing, namely the generalized smoothing approach  \citep[GSA,][]{ramsay2007parameter} and the manifold-constrained Gaussian processes \citep[MAGI,][]{yang2021inference}. For these comparisons, we use harmonic equations for modelling a two-dimensional motion \citep{ramsay2017dynamic}
\begin{eqnarray*}
    D^2X(t)&=c_{X}X(t) + a_X(t),\\
    D^2Z(t)&=c_{Z}Z(t) + a_Z(t),
\end{eqnarray*}
where $t\in [0,6]$, and $X(t)$ and $Z(t)$ represent the locations of an object in the horizontal and vertical directions at time $t$, respectively. The scalar parameters $c_X$ and $c_Z$ are involved, along with the forcing terms $a_X(t)$ and $a_Z(t)$ promoting the motion. We assume that $a_X(t)$ and $a_Z(t)$ are both step functions on the interval $[0,6]$ with $Q$ equal segments. The parameters $c_X$, $c_Z$, and the step functions $a_X(t)$ and $a_Z(t)$ are the unknown identities of interest, containing in total $(2Q+2)$ parameters to be estimated from the contaminated data of $X(\cdot)$ and $Z(\cdot)$. 

\begin{table}[h]
\centering
\renewcommand{\arraystretch}{1.1}
\footnotesize
\caption{Average runtimes and RRMSEs of five methods based on 100 simulations on a computer with 2.20 GHz Intel(R) Xeon(R) CPU and 128 GB RAM.}
\begin{tabular}{c|c|c|c|c|c|c|c}
\hline
\multicolumn{2}{c|}{ }   &  \multicolumn{3}{c|}{Runtime (sec)} &  \multicolumn{3}{c}{RRMSE (\%)}\\
   \cline{3-8}
\multicolumn{2}{c|}{ }
   &  \multicolumn{1}{c|}{$\gamma=3\%$} &  \multicolumn{1}{c|}{$\gamma=5\%$}& \multicolumn{1}{c|}{$\gamma=7\%$}  & \multicolumn{1}{c|}{$\gamma=3\%$} &  \multicolumn{1}{c|}{$\gamma=5\%$}& \multicolumn{1}{c}{$\gamma=7\%$}\\
   \hline
    \multirow{5}{*}{$n=50$} 
&Order 2 gradient matching & 2.02 & 1.97 & 1.92 & 11.34 & 16.35 & 20.64 \\ 
   & Order 1 gradient matching & 2.19 & 2.14 & 2.08& 10.99 & 15.34 & 18.03 \\ 
   & Green's matching & 2.70 & 2.66 & 2.60& 2.64 & 4.25 & 5.88 \\ 
   & GSA & 3.14 & 3.30 & 3.24& 8.39 & 9.36 & 10.48  \\ 
   & MAGI & 217.96 & 227.48 & 223.23  & 12.43 & 7.99 & 9.58 \\
   \cline{1-8}
   \multirow{5}{*}{$n=150$} 
& Order 2 gradient matching & 5.86 & 5.09 & 4.96 & 10.65 & 13.38 & 16.06  \\ 
   & Order 1 gradient matching & 5.99 & 5.21 & 5.09 & 10.21 & 12.69 & 14.72  \\ 
   & Green's matching & 6.54 & 5.77 & 5.62 & 1.60 & 2.60 & 3.61  \\ 
   & GSA &  43.59 & 40.46 & 37.19 & 1.84 & 2.69 & 3.60 \\ 
   & MAGI & 892.46 & 788.41 & 742.93&  3.83 & 4.93 & 5.87  \\ 
     \cline{1-8}
    \multirow{5}{*}{$n=250$} 
& Order 2 gradient matching & 9.10 & 9.08 & 8.49 & 10.02 & 12.58 & 14.69 \\ 
   & Order 1 gradient matching & 9.20 & 9.20 & 8.60 & 8.94 & 11.60 & 13.30 \\ 
   & Green's matching & 9.71 & 9.69 & 9.12 & 1.25 & 1.97 & 2.70 \\ 
   & GSA & 192.91 & 184.99 & 169.05 &  1.35 & 1.97 & 2.63 \\ 
   & MAGI & 2475.61 & 2347.36 & 2187.9  & 3.01 & 3.84 & 4.66\\  
   
\hline
\end{tabular}\label{ode_resultt}
\end{table} 

To implement GSA and MAGI, we first transform the above equations into  first-order equations by defining new system trajectories $x(t):=DX(t)$ and $z(t):=DZ(t)$. After that, GSA or MAGI employs B-spline basis functions or Gaussian processes to model $X(\cdot)$, $Z(\cdot)$, $x(\cdot)$, and $z(\cdot)$, and estimate the parameters as well as $X(\cdot)$, $Z(\cdot)$, $x(\cdot)$, and $z(\cdot)$ under a penalized likelihood or Bayesian framework. 
Typically, these two methods both treat the dynamic systems as equation constrains on some evaluated points from $[0,C]$, similar to evaluating Green's matching on $\{t_h\in [0,C];h=1,\cdots, H\}$ in Section \ref{op_gree_1}. We abuse the notation $H$ to denote the number of evaluated points for GSA and MAGI. 

In general, GSA employs an iterative optimization procedure for the parameter estimation, in which one iteration requires $O((n+H)B^2+B^3)$ computations, with $B$ being the number of B-spline bases used. Besides, MAGI requires $O(H^3)$ computations to factorize covariance matrices of Gaussian processes within each Bayesian sampling. The computational costs of these two methods are significantly higher than those of the two-step methods, which only need to solve least squares regressions after the pre-smoothing step. The computational complexity of the pre-smoothing is generally $O(n+H)$ when $\{t_h\in [0,C]; h=1,\cdots, H\}$ is equally spaced \citep{fan2018local}, with $O(H)$ computations for the least squares fitting.

During our numerical analysis, we notice that the results of GSA and MAGI are sensitive to the initial parameters provided for the optimization or sampling. To avoid computationally intensive initial value searching, we use the estimated parameters obtained by Green's matching as the initial values of parameters for both GSA and MAGI.
Additionally, as we obtain posterior samples of parameters in MAGI, we calculate the average of the 1500 posterior samples as the point estimates for MAGI.

We conduct additional simulation studies based on the harmonic equations following the settings described in the last subsection. We assumed that $c_X=c_Z=-0.0015$, $\alpha_X(t)$ and $\alpha_Z(t)$ are monotone step functions varying from -0.0005 to 0.0005 and from -0.002 to 0.001 with $Q=6$, respectively, and $X(0)=Z(0)=DX(0)=DZ(0)=0$. The average runtimes and the RRMSEs of the five methods are presented in Table \ref{ode_resultt}.

Based on Table \ref{ode_resultt}, we observe that the runtimes of the two-step methods are significantly smaller than those of the other two methods, especially for the cases of $n=150$ or $250$. With the computational convenience, Green's matching outperforms gradient matching of different orders and MAGI, and is superior to GSA when the sample size $n$ is relatively small.
Although GSA performs slightly better than Green's matching for the cases of $n=150$ and $\gamma=7\%$ or $n=250$ and $\gamma=7\%$, the runtimes of GSA are substantially larger than those of Green's matching in these cases.

To further illustrate different methods, we apply the previous harmonic equations to model a handwriting data in Part 4 of Supplementary Materials.
The data illustration again demonstrates the competitive performance of Green’s matching.
Overall, considering both computational and statistical efficiency, Green's matching is an appropriate choice for parameter estimation in harmonic equations.

\section{Equation discovery of a nonlinear pendulum}\label{ED}
In this section, we apply Green's matching to discover the governing equation of a nonlinear pendulum. The dynamic system is given as follows
\begin{eqnarray*}
    D^2X(t)=-\sin(X(t)).
\end{eqnarray*}
In this task, we do not assume a known form of the right-hand side in the above equation, and aim to infer this unknown part using the contaminated data of $X(\cdot)$.
In \citet{brunton2016discovering}, this tack is facilitated by embedding sparsity-pursuit regression methods, such as Lasso \citep{hastie2009elements}, into the gradient matching.
Here, we can generalize this idea to Green's matching.
We assume that 
\begin{eqnarray}\label{NP}
    D^2X(t)=\bm{g}^T(X(t))\ \bm{\beta},
\end{eqnarray}
where $\bm{g}(\cdot)$ is a $\mathbb{R}^q$-valued known candidate function and $\bm{\beta}=(\beta_1,\cdots,\beta_q)^T$ are the unknown parameters to be estimated from data. Following \citet{brunton2016discovering}, we set that $\bm{g}(X(t))=(1,X(t),X^2(t), X^3(t), X^4(t),\sin(X(t)),\cos(X(t)))^T$, containing the constant, polynomial, and trigonometric functions of $X(t)$. Under this framework, we modify the Green's matching \eqref{gre_11} as
\begin{eqnarray*}   \min_{\bm{{\alpha}}\in\Omega_{\bm{{\alpha}}},\bm{\beta}\in\Omega_{\bm{\beta}}}\int_{0}^C\sum_{i=1}^p\left[\mathcal{T}^{i,\text{Gree}}_{t,\bm{\beta},\bm{\alpha}_i}(\bm{\hat{X}})\right]^2w(t)\ \mathrm{d}t+\lambda\sum_{l=1}^q|\beta_l|,
\end{eqnarray*}
where $\lambda$ is a tuning parameter to introduce sparsity for $\bm{\beta}$, allowing identifications of the true functional relations from the observed data. Since \eqref{NP} satisfies the separability \eqref{de_f}, the above optimization is essentially a sparse least squares problem. Therefore, it can be efficiently solved by many existing methods.

\begin{figure}[h]
\centering
\includegraphics[scale = 0.6]{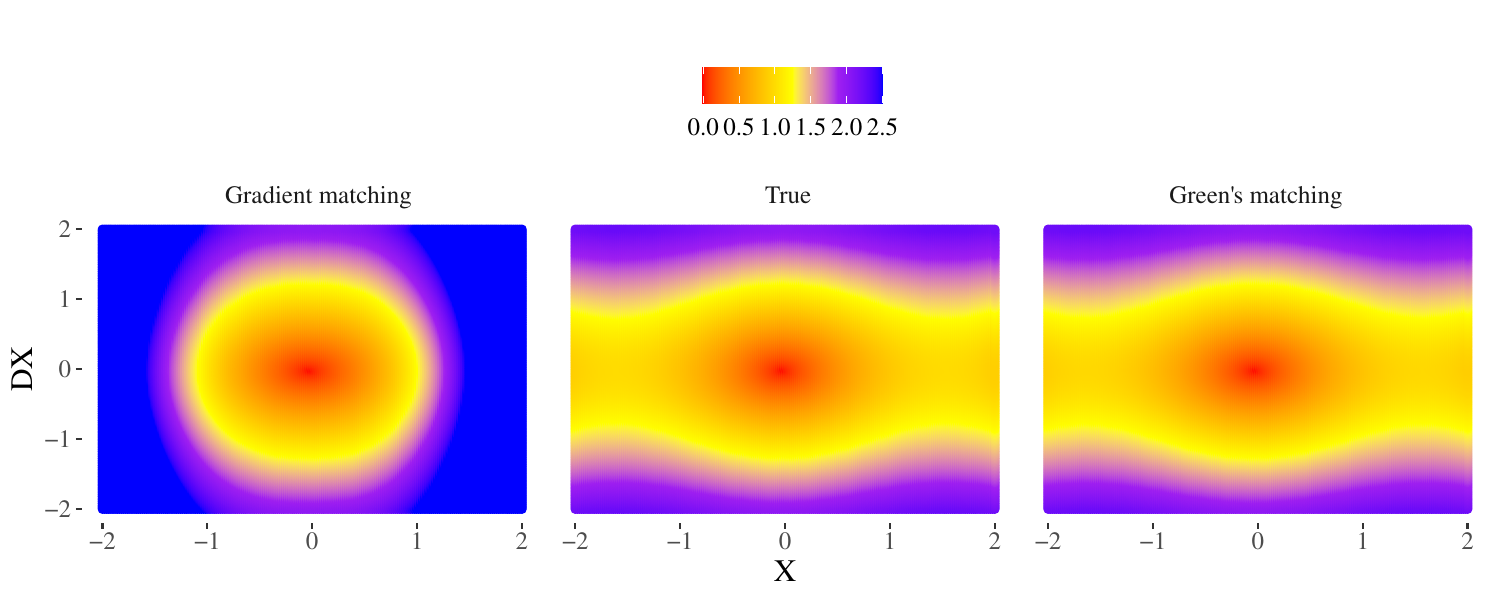}
\caption{The true vector field and the average vector fields  from 100 simulations for the nonlinear pendulum, where the color represents the magnitude of the Euclidean norm of $(DX(t),D^2X(t))^T$.}\label{EDNP}
\end{figure}

We generate 100 replications of simulated data for the nonlinear pendulum by setting $n=50$ and $\gamma=5\%$, in which the initial values $(X(0),DX(0))$ are randomly sampled from the uniform distribution on $[-0.5,0.5]$. After that, we respectively apply gradient matching and Green's matching for discovering equations from the observed data. It's worth noting that the discovered equation determines $(DX(t),D^2X(t))$ given $(X(t),DX(t))$, i.e., determining a vector field from $\mathbb{R}^2$ to $\mathbb{R}^2$. To compare different methods, we present the true and the estimated vector fields for the nonlinear pendulum in Figure \ref{EDNP}. From these results, we observe that the estimated vector fields by Green's matching are nearly the same as the true case, whereas those obtained by gradient matching exhibit significant biases outside the central region. These biases may arise from the approximation of high-order derivatives in the gradient matching, prohibiting its effectiveness in discovering equations from discrete noisy data. 
In contrast, Green's matching demonstrates its superiority in equation discovery by only approximating $X(\cdot)$ for the estimation.

\section{Discussion}\label{dis}

\subsection{Advantages and limitations}

This article presents an efficient two-step method called Green's matching for parameter estimation in general-order dynamic systems. Our method is simple to implement with the aid of pre-smoothing under the equation matching framework.
By a transformation induced by Green's function, we avoid the need to approximate curves' derivatives within the parameter estimation. This can reduce the smoothness conditions required for the parameter estimation, a significant advantage of Green's matching compared to gradient matching, integral matching, and other competitive methods in the literature \citep{ramsay2007parameter, yang2021inference}.
We prove that Green's matching obtains the statistical optimal root-$n$ consistency for general-order dynamic systems. 
However, the other two-step approaches may not achieve this optimal rate when the differential order is relatively high.
The simulation studies, data illustration, and the task of equation discovery also demonstrate the superiority of Green's matching over the existing methods.

It should be noted that Green's matching needs a satisfactory pre-smoothing of the trajectories $\bm{X}$ for the parameter estimation. Therefore, our method generally prefers that the observed trajectories are densely observed (i.e., $n$ tends to infinity), and may not be suitable for sparsely observed trajectories. 
This is one limitation of Green's matching compared to the method without pre-smoothing, such as \citet{ramsay2007parameter}.
Nonetheless, we can similarly generalize the idea from \citet{ramsay2007parameter} to Green's matching for sparsely observed curves. 
Intuitively, this can be facilitated by borrowing information of the loss function \eqref{gre_22} for smoothing $\bm{X}$, thereby enhancing the statistical efficiency for approximating $\bm{X}$ when $n$ is relatively small.

\subsection{Extensions for complex tasks}
For parameter estimation of general-order systems, we primarily focus on cases with a fixed $K$ in this article. This is reasonable for many real-world applications, as $K$ can usually be specified based on prior knowledge of dynamic systems. However, in more general scenarios where the differential order is not explicitly given, it could be valuable to determine $K$ from the data due to its importance for constructing Green's function and implementing Green's matching. In such cases, we can treat $K$ as a tuning parameter to be estimated by some data-driven criterion. This is one potential extension of our method to infer the differential order of dynamic systems.

Currently, more complicated statistical inferences in dynamic systems have also aroused great interest in various domains, e.g., discovering laws of equations \citep{brunton2016discovering,champion2019data} or inferring causal networks \citep{wu2014sparse,chen2017network,zhang2020bayesian,dai2021kernel}. A unified way to handle these problems is to assume a flexible representation of $f_i(\cdot;\cdot)$ in \eqref{dyn} similar to the model in Section \ref{ED}. By plugging $\bm{X}(\cdot)$ and its derivatives (or their approximations) into some equations, inferring $f_i(\cdot;\cdot)$ can be viewed as a common machine learning problem, and hence many tools of machine learning can be directly applied in this circumstance.
Among these applications, gradient matching and integral matching are two versatile frameworks to introduce the first-step approximations of curves for dynamic systems. However, as we have demonstrated, these two frameworks may lead to unreliable parameter estimations with large biases for the second-order (or higher-order) systems. Targeting these complex systems, Green's matching provides a universal framework with fewer smoothness assumptions for statistical inferences. This appealing feature suggests that Green's matching may be more suitable than the other two methods for discovering acceleration-based laws of equations (e.g., the physical equations based on Newton's second law) or inferring causal networks among a collection of second-order oscillatory trajectories.
We leave these as the future directions we plan to investigate.

\bibliographystyle{apalike}
\bibliography{refbib}

\end{document}